\documentclass[twocolumn]{openjournal}

\usepackage{graphicx}	
\usepackage{amsmath}	
\usepackage{amssymb}	
\usepackage{soul}

\usepackage{booktabs}

\usepackage{xcolor}

\usepackage{float}
\usepackage{fontawesome}
\usepackage[colorlinks,allcolors=blue]{hyperref}
\usepackage{url}

\usepackage{listings}
\usepackage{array,multirow,graphicx}

\definecolor{maroon}{cmyk}{0, 0.87, 0.68, 0.32}
\definecolor{halfgray}{gray}{0.55}
\definecolor{ipython_frame}{RGB}{207, 207, 207}
\definecolor{ipython_bg}{RGB}{247, 247, 247}
\definecolor{ipython_red}{RGB}{186, 33, 33}
\definecolor{ipython_green}{RGB}{0, 128, 0}
\definecolor{ipython_cyan}{RGB}{64, 128, 128}
\definecolor{ipython_purple}{RGB}{170, 34, 255}

\begin{document}

\journalinfo{The Open Journal of Astrophysics}
\submitted{submitted XXX; accepted YYY}

\title{Probing Environmental Dependence of High-Redshift Galaxy Properties with the Marked Correlation Function}
\author{Emy Mons, Charles Jose}
\affiliation{Department of Physics, CUSAT, Cochin, 682022, India}
\thanks{$^\star$ E-mail: \href{emymons92@gmail.com}{emymons92@gmail.com},
\href{charles.jose@cusat.ac.in }{charles.jose@cusat.ac.in }}

\begin{abstract}

In hierarchical structure formation, correlations between galaxy properties and their environments reveal important clues about galaxy evolution, emphasizing the importance of measuring these relationships. We probe the environmental dependence of Lyman-break galaxy (LBG) properties in the redshift range of $3$ to $5$ using marked correlation function statistics with galaxy samples from the Hyper Suprime-Cam Subaru Strategic Program and the Canada–France–Hawaii Telescope U-band surveys.
We find that the UV magnitude and color of magnitude-selected LBG samples are strongly correlated with their environment, making these properties effective tracers of it. In contrast, the star formation rate and stellar mass of LBGs exhibit a weak environmental dependence. For UV magnitudes and color, the correlation is stronger in brighter galaxy samples across all redshifts and extends to scales far beyond the size of typical dark matter halos. This suggests that within a given sample, LBGs with high UV magnitudes or colors are more likely to form pairs at these scales than predicted by the two-point angular correlation function. Moreover, the amplitude of the marked correlation function is generally higher for LBG samples compared to that of $z \sim 0$ galaxies from previous studies. We also find that for LBG samples selected by the same absolute threshold magnitude or average halo mass, the correlation between UV magnitudes and the environment generally becomes more pronounced as the redshift decreases. On the other hand, for samples with the same effective large-scale bias at $z\sim 4$ and $5$, the marked correlation functions are similar on large scales.

\end{abstract}

\maketitle



\section{Introduction}
In the standard paradigm of cosmological structure formation, galaxy formation and evolution are heavily influenced by their surroundings, leading to a strong correlation between galaxy properties and environment. Observational probes of these correlations have emerged as a robust tool for constraining galaxy formation models.
The galaxy clustering, measured by the two-point correlation functions, has been successfully used for this purpose; these studies firmly established that for $z \lesssim 1$, galaxies that are redder, passively evolving, and early-type exhibit stronger clustering, indicating they are situated in regions of high dark matter density. Conversely, bluer, star-forming, and late-type galaxies reside in less dense environments \citep{Norberg2002, Zehavi2005, Coli_Newman+2008_color_xir_deep2, zehavi+2011, Guo_zehavi+2013_xir_color_sdss, LawSmith_Eisenstein_2017_xir_depend, Lin_Fang+2019_color_xir_cosmos}. 
Other clustering studies positively confirm the observed link between environment and galaxy properties like luminosity, stellar mass ($M_\ast$), age, morphology, and spectroscopic features \citep{Mostek+2013_xir_depend_deep2, Cook+2013_LBG_env_2pcf, Cochrane+2018_xir_depend, Durkalec+2018_xir_depend_vimos,  Zhai_Percival_2023_xir_depend_boss}. 

The impact of the environment on galaxy properties is also examined using the local density of neighboring galaxies surrounding each galaxy using nearest-neighbor methods that determine the volume containing a fixed number of nearest-neighbors or fixed-aperture measurements, which estimate galaxy number density within a given volume surrounding the galaxy \citep[see][for details]{muldrew_2012_gal_env}. These methods have been widely applied to vast datasets from galaxy surveys, particularly at lower redshifts, for analyzing the correlations between galaxies and their environments. Studies have found that for $z \lesssim 1$, the star formation in galaxies within over-dense regions (of high matter or galaxy number density) is quenched \citep{Kauffmann+2004, Cooper+2008, Peng+2012, woo_dekel+2013, Koyama+2018_sfr_env_subaru, old_balogh+2020, Gu+2021_env_SFR_candels} and the quenching efficiency is dependent on parameters like the stellar mass of the galaxy and increases over time \citep{Sobral+2011_sfr_env_Mstar, darvish+2016_sfr_quench_z1, chartab+2020_SFR_env_candels, Mao_Kodama+2022_galaxy_quench_sfr_env_z1_cosmos}. 

Unlike the well-studied low-z Universe, there are comparatively fewer studies at high redshifts ($z \geq 2$) that explore the environmental dependence of galaxy properties. Furthermore, the findings from these high-redshift studies often do not agree with each other. For instance, there are studies that suggest that galaxies in dense regions exhibit higher star formation rate (SFR) compared to those in low-density regions \citep{shimakawa+2018_sfr_env, Lemaux+2022, Taamoli+2023, shi+2024_sfr_env_z4}. However, other observational and theoretical studies report either no significant correlation \citep{Darvish+2015_galaxy_env_z3} or even an anticorrelation \citep{chartab+2020_SFR_env_candels, Hartzenberg+2023_sfr_quench_z2.5} between SFR and the surrounding galaxy overdensity.  

Moreover, there are even fewer studies at high redshifts that investigate the correlation between the environment and other properties such as luminosity/magnitude or color. Given these, it is crucial to understand how different galaxy properties—like luminosity, color, $M_\ast$, and SFR—trace the galaxy environment at high redshifts. In this work, we address this question using marked correlation function (MCF) statistics \citep{beisbart+2000}, a powerful tool for detecting and quantifying the correlation between galaxy properties and their environment.

The MCF is a two-point statistic that, unlike the conventional two-point correlation function, assigns a weight/mark to each galaxy based on the value of its physical properties. The resulting clustering relative to the unmarked clustering is termed the MCF and is a strong indicative of the environmental dependence of properties of galaxies. Previous theoretical and numerical investigations have shown that MCF is a robust tool for tracing galaxy environments and galactic conformity and is complementary to the aforementioned observables \citep{sheth2004environmental, Sheth2005_marked, sheth2006environment, harker2006marked, skibba2013measures, zu_2018_conformity_mcf, Xiao2022}. It encapsulates richer information compared to numerous other environmental probes because it necessitates the values of galaxy properties (e.g., luminosities) rather than relying on binary classification (such as bright or faint). The MCF has also proven effective in detecting subtle environmental correlations \citep{CalderonV_BerlindA+2018}, and easier to model analytically in the framework of the halo model of large-scale structures \citep{Sheth2005_marked, skibba2013measures}.

Several recent works use the MCF as an effective tool for probing the environmental dependence of galaxy properties like $M_\ast$, SFR, luminosity/magnitude in different bands, and colors at $z \sim 0$ \citep{rutherford2021sami, Riggs2021, sureshkumar2021_gamaMCF, sureshkumar2023_midinfra, sureshkumar2024_merger}. In particular, \citet{sureshkumar2021_gamaMCF, sureshkumar2023_midinfra}, using  Galaxy and Mass Assembly (GAMA) the Wide-field Infrared Survey Explorer (WISE) survey observations, showed that K-band luminosity and $M_\ast$ are the most direct tracer of the galactic environment in the local Universe. 

In this paper, we advance these studies to very high-z by investigating the environmental dependence of Lyman Break Galaxies (LBGs) using MCF in a broad redshift range of $3 \leq z \leq 5$. For this, we utilize the wide, deep and ultra-deep layers of the latest Hyper Suprime-Cam Subaru Strategic Program (HSC-SSP) data release and the deep and ultra-deep U-band imaging data provided by the Canada-France-Hawaii Telescope Legacy (CFHT) U-band Deep Survey \citep{aihara+2022_pdr3_subaru, sawicki+2019_clauds}. The multi wavelength data set allows us to identify sufficiently large samples of both faint and bright galaxies to measure the MCFs fairly accurately and draw conclusions on how different properties of high-z galaxies are correlated with their environment.

Our work complements other very high-z probes of environmental dependence of LBG properties using alternative statistics \citep{Lemaux+2022, shi+2024_sfr_env_z4, Taamoli+2023}. These studies investigate correlations of galaxy properties from ultra-deep surveys (of survey area $\lesssim 2$ square degrees) including CANDELS \citep{Grogin+2011_candels}, VIMOS \citep{LeFevre+2025_VIMOS}, and COSMOS \citep{weaver+2022_cosmos}. 
Our study, however, leverages data from the wide and deep surveys (of survey area $\sim$ 600 square degrees) of HSC-SSP and CFHTU to investigate, with robust statistical significance, the correlations in LBG samples that are up to two orders of magnitude brighter than those used in previous studies. We further extend our analysis to higher redshifts than those explored in these works. In addition, we compare the marked clustering strength of galaxy samples with similar absolute magnitude, effective large-scale bias, and effective halo mass across different redshifts to probe how these factors shape the evolution of environmental dependencies of galaxy properties with redshift.

The structure of this paper is as follows: Section~\ref{sec_data} details the data, while Section~\ref{sec_mcf} describes the methodology used to measure the MCF of LBGs. The measurements of MCFs for various galaxy samples at different redshifts and their implications are discussed in section~\ref{sec_result}. We provide the summary of the results in the section~\ref{sec_summary}. The cosmological parameters used for the calculation of the effective large-scale bias and the effective halo mass in sections~\ref{sec_same_bias} and \ref{sec_same_mass} are adopted from \citet{Planck_2018}.  

\section{The Date}
\label{sec_data}
\subsection{The HSC-SSP data}
\label{sec_hsc_ssp_data}
We use publicly available galaxy catalogues from the 3$^\mathrm{rd}$ public data release (PDR3) of the HSC-SSP survey \citep[][see also \citealt{miyazaki2018hyper,aihara+2018_pdr1_subaru, aihara2019second}]{aihara+2022_pdr3_subaru}. The HSC-SSP comprises of three survey layers of different depths: wide, deep, and ultra-deep, with area coverages of about 670, 28, and 2.8 square degrees in five optical broad-band filters ($g$, $r$, $i$, $z$, and $y$) \citep{kawanomoto2018hyper} (see fig.~1 of \citep{aihara+2022_pdr3_subaru} for details of the area covered). In this work, we use galaxies from the Spring and Autumn regions of the wide survey and the COSMOS, DEEP2-3, ELAIS-N1, and XMM-LSS regions of the deep survey. The galaxy catalogues are obtained by reducing the  HSC data using the \textsc{hscPipe},  the HSC data reduction pipeline \citep{Bosch+2018_hscpipe,huang2018characterization}. These catalogues are used to probe the clustering properties of magnitude-limited samples of Lyman Break Galaxies (LBGs) at $z \sim 4$ (g-dropouts) and $z \sim 5$ (r-dropouts), selected using the Lyman break color technique \citep{steidel1996spectroscopic, giavalisco2002lyman} that traces the redshifted Lyman-limit emission from the galaxy at the rest-frame wavelength $\lambda = 912 \AA $.  

As in \citet{harikane2022goldrush}, we select galaxies with signal-to-noise ratios higher than $5$ within $1.5''$ diameter apertures and apply the masks \citep{mandelbaum2018first,li2022three}, threshold values, and flags summarized in section 2 and table 2 of their paper, to remove galaxies affected by pixel issues, cosmic rays, and bright source halos \citep{coupon2018bright,furusawa2018site}. The color selection criteria for the dropout galaxies are applied to convolvedﬂux\textunderscore0\textunderscore20, the $2''$ diameter apertures magnitudes after aperture correction, which for g-dropouts (z $\sim$ 4) is \citep{ono2018great, toshikawa2018goldrush, harikane2018goldrush} 
    \begin{equation}
    ( g - r > 1.0 ) \Lambda  ( r - i < 1.0 ) \Lambda  ( g - r > 1.5 ( r - i ) + 0.8 ).  
    \end{equation}
The selection criteria for r-dropouts (z $\sim$ 5) is given as
    \begin{equation}
    ( r - i > 1.2 ) \Lambda ( i - z < 0.7 ) \Lambda ( r - i > 1.5 ( i - z ) + 1 ).  
    \end{equation}

Once the galaxies are color-selected, the ﬁxed $2''$ diameter aperture magnitude after aperture correction is used as the total magnitude in all filters \citep{harikane2022goldrush}. The UV magnitude ($\text m_{\text UV}$) for the g- and r-dropouts are, respectively, the i- and z- band magnitudes, whose central wavelengths are the nearest to the rest frame wavelength of $\lambda = 1500 \AA$. 

We also use the photometric redshifts (photo-z) of the galaxies as given by PDR3, derived using the empirical photo-z fitting code DEmP \citep{hsieh2014estimating, tanaka2018photometric}. 
The DEmP utilizes template data of galaxies with spectroscopic redshifts ($z_{\text{sp}}$) from the literature to obtain an accurate polynomial fit to the probability distribution function (PDF) of photo-z of galaxies  as a function of a small number multi-broadband magnitudes  \citep{hsieh+2014_demp}. The PDF is then used to derive various point estimates, confidence intervals, and percentiles of the photo-z ($z_{\text{ph}}$). We use the best-fit photo-z derived using the DEmP due to their lower outlier rate at all magnitudes. The bias in the estimated photo-z is within $1\%$ across all magnitudes and redshifts. The scatter in photo-z, given by the standard deviation of $\delta z = (z_{\text{ph}} - z_{\text{sp}})/(1 + z_{\text{sp}})$, generally increases as the magnitudes become fainter, varying between $0.15$ and $0.1$ in the redshift range of our study \citep{tanaka2018photometric}. In addition to photo-z, we adopt the mode values of the $M_\ast$ and SFR computed from respective PDFs. 

\subsection{CLAUDS HSC-SSP data}
The CFHT U-band Survey (CLAUDS) \citep{sawicki+2019_clauds} in the deep layer of the HSC-SSP program is useful for selecting $z\sim3$ LBGs via the U-dropout technique. The U-band images are obtained with two ﬁlters, u or u$^\ast$, in four deep fields:  COSMOS, DEEP2-3, ELAINS-N1, and XMM-LSS, covering about 20 square degrees. The COSMOS and XMM-LSS regions contain ultra-deep fields with each of area $0.8$ square degrees. We utilize the combined galaxy catalogue from \citet{desprez+2023_clauds}, which is based on CLAUDS data processed with SExtractor \citep{bertin1996sextractor} and HSC-SSP data reduced using \textsc{hscpipe}. From the catalogue, we have selected clean galaxies by applying suitable flags and masks as described in \citet{desprez+2023_clauds}. The  color selection criteria for U-dropouts ($z \sim 3$) is described in section 3.5 of \citep{harikane2022goldrush} and is given by
    \begin{equation}
    ( g - r < 1.2 ) \Lambda ( u^\ast - g > 0.9 ) \Lambda  ( u^\ast - g > 1.5 ( g - r ) + 0.75 )    \nonumber
    \end{equation}
   \begin{center}
       or
   \end{center}
     \begin{equation}
      ( g - r < 1.2 ) \Lambda ( u - g > 0.98 ) \Lambda  ( u - g > 1.99 ( g - r ) + 0.68 ). 
    \end{equation}

The catalogue also provides the source redshifts estimated using two template fitting codes: (i) \textsc{Le Phare} for SExtractor catalogue \citep{Arnouts+2011_lephare} and (ii) \textsc{Phosphoros} for \textsc{hscpipe} catalogue (Paltani et al., in prep.; \citet{desprez+2020_euclid}).The UV magnitude ($\text m_{\text UV}$) for the U-dropouts is the r-band magnitude. We additionally adopt the best-fit values of $M_\ast$, and SFR provided in the catalogue for the ultra-deep regions; however, these values are not available for the deep region. Table~\ref{table1} summarizes the number of dropout galaxies in the samples analyzed in this study. 

\subsection{Removal of bright spurious sources and low-z interlopers}

The presence of extremely bright spurious stellar sources in the data could introduce potential biases in the clustering measurements. To address this, we applied a bright magnitude cutoff of $\text{m}_\text{UV} \geq 20$ as in \citet{harikane2022goldrush} at every redshift to exclude such sources. Adjusting this cutoff to nearby values has minimal impact on the sample size as spurious sources are much brighter.  For instance, at $z\sim 3$ lowering the cutoff to $18$ adds only 1 source, while changing it to $\text{m}_\text{UV} \geq 22$ removes 33 sources compared to the sample with the cutoff at $20$.

Low-z interlopers satisfying the above color-selection criteria contaminate the HSC-SSP and CLAUDS samples. To remove these, as in \citet{toshikawa+2024}, we first select galaxies with the upper bound on the 95 percent confidence interval of the photo-z ($z_{95}$) greater than $2.8$ for g-dropouts and $3.8$ for r-dropouts. For U-dropouts, the upper bound on the 68 percent confidence interval of the photo-z is taken to be greater than $2.3$, and galaxies are further selected with  $2.6 \leq z \leq 3.4$. This constitutes our final sample, which will be used for clustering analysis. 

\begin{figure}
    \centering
    \includegraphics[width=1\linewidth]{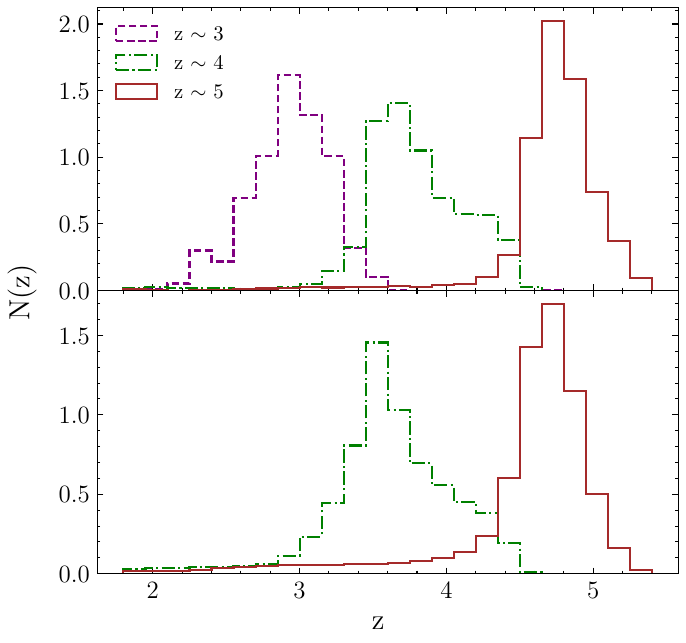}
    \caption{Top panel: Normalized redshift distribution of U-, g-, and r-dropouts from  CLAUDS/HSC-SSP deep survey. Bottom panel: The redshift distribution of g-, and r-dropouts from the wide survey of HSC-SSP.}
    \label{fig_Nz}
\end{figure}

In Fig.~\ref{fig_Nz}, we show the photo-z distribution, $N(z)$, of the U-, g-, and r-dropouts from the deep and wide surveys. For g- and r-dropouts, a Kolmogorov–Smirnov test was conducted to assess the differences in $N(z)$ between the wide and deep datasets at the same redshift. The resulting p-value was close to $0$, implying that the redshift distributions of the deep and wide datasets differ at each redshift. We note that the angular correlation function is quite sensitive to the $N(z)$ of the galaxy sample.  However, as the marked correlation function (MCF) is defined as the ratio of weighted to unweighted angular correlation functions (see Section~\ref{sec_mcf}), it is expected to be less affected by $N(z)$.

\begin{table}
\caption{The number of U-, g-, and r-dropouts at various magnitude thresholds used in our study.}
\begin{tabular}{c c c c}
\hline  \\[0.01cm] 
Dropouts & Region & \begin{tabular}[c]{@{}c@{}}Limiting Magnitude\\ (m$_\text{UV}$)\end{tabular} & \begin{tabular}[c]{@{}c@{}}Number of \\ galaxies \end{tabular} \\ \hline
\hline \\[0.01cm] 
\multicolumn{1}{c}{\multirow{6}{*}{U }} & \multirow{4}{*}{deep} & 24 & 4151 \\ [1ex]
\multicolumn{1}{c}{} &  & 24.5 & 14,329\\ [1ex] 
\multicolumn{1}{c}{} &  & 25 & 38,801 \\ [1ex]  
\multicolumn{1}{c}{} &  & 25.5 & 82,942 \\ [1ex]  \cline{2-4} \\[0.01cm] 
\multicolumn{1}{c}{} & \multirow{2}{*}{ultra deep} & 26 & 20,843  \\ [1ex]  
\multicolumn{1}{c}{} &  & 26.5 & 34,019 \\ [1ex] \hline \\ [0.01cm]
\multicolumn{1}{c}{\multirow{4}{*}{g }} & \multirow{2}{*}{wide} & 24 & 28,145 \\ [1ex]
\multicolumn{1}{c}{} &  & 24.5 &1,58,305 \\ [1ex] \cline{2-4} \\[0.01cm] 
\multicolumn{1}{c}{} & \multirow{2}{*}{deep} & 25 & 14,194 \\ [1ex]
\multicolumn{1}{c}{} &  & 25.5 & 38,408 \\ [1ex] \hline \\[0.01cm]
\multicolumn{1}{c}{\multirow{3}{*}{r }} & \multirow{1}{*}{wide} & 24.5 &  38,633 \\ [1ex]
\cline{2-4} \\[0.01cm] 
\multicolumn{1}{c}{} & \multirow{2}{*}{deep} & 25 & 3828  \\[1ex]
\multicolumn{1}{c}{} &  & 25.5 & 10,768 \\ [1ex] \hline
\end{tabular}

\label{table1}
\end{table} 
\section{The MCF }
\label{sec_mcf}
In this section, we describe the MCF statistics, which will be used to probe the environmental correlations of properties of the galaxy samples. The MCF measures how the properties of galaxy pairs—rather than individual galaxies—depend on their separation. Specifically, it is the correlation coefficient of the properties of galaxies (such as luminosity, color, $M_\ast$, SFR, morphology, etc.) relative to the standard unweighted correlation function. The MCF provides additional information compared to traditional fixed-aperture measurements and the nearest-neighbor statistic \citep{CalderonV_BerlindA+2018} and has the potential to place strong constraints on cosmological parameters \citep{Yang_2020_cosmology, Xiao2022_mcf_cosmology, LaiDing2023, Massara2023_mcf_cosmology}, modified gravity models \citep{white2016_mcf_modgrav, satpathy2019measurement, Aviles2021, Aviles2020, Hernandez2018, Armijo2018}, and neutrino mass \citep{Massara2021_mcf_neutrino}.

Before computing the MCF, it is useful to define the standard two-point correlation function (2PCF), which represents the excess probability of finding two galaxies separated by a given spatial or angular scale compared to what is expected in a Poisson point process \citep{Peebles1980}. The angular 2PCF is estimated using the well-known Landy-Szalay estimator \citep{landy1993bias} as
\begin{equation}
            \omega (\theta)= \frac{\langle DD(\theta)\rangle-\langle 2DR(\theta)\rangle+\langle RR(\theta)\rangle}{\langle RR(\theta)\rangle}\, , 
            \label{eq_wtheta}
\end{equation}
where $\langle DD(\theta) \rangle$, $\langle DR(\theta)\rangle$, and $\langle RR(\theta)\rangle$ are the numbers of galaxy-galaxy, galaxy-random, and random-random pairs at a pair separation angle $\theta$, normalized by the total number of pairs. A random catalogue that accounts for masked areas of the survey region and edge effects is necessary to compute the random pairs. In this work, we use the random catalogues provided by \cite{aihara+2022_pdr3_subaru} (see also \citet{coupon2018bright}), which have a surface density of 100 points per square arcminute.

The MCF is defined as in \citet{skibba2013measures} \citep[see also][]{stoyan1994fractals, beisbart+2000, Sheth+2004_marked, Sheth2005_marked, Sheth+2006_marked, skibba2006MCF, martinez2010measuring,sureshkumar2023}.
 \begin{equation} 
    M(\theta)= \dfrac{1+W(\theta)}{1+\omega (\theta)} \equiv \dfrac{WW(\theta)}{DD(\theta)} \,, 
    \label{eq_mcf}
\end{equation}
where W($\theta$) is the weighted angular correlation function, estimated similarly to the $\omega(\theta)$ but with an additional step of assigning a weight or marks to each galaxy based on its property value. 

To compute the MCF,instead of explicitely calculating $W(\theta)$ and $\omega(\theta)$, we use the estimator $WW(\theta)/DD(\theta)$ on the RHS of Eq.~\ref{eq_mcf}. Here, $WW(\theta)$ are weighted pair counts computed after weighing each galaxy with a mark.   
\begin{equation}
    WW(\theta)= \sum_{ij} w_i w_j \,, 
\end{equation}
where $w_i$ is the mark of the $i^\text{th}$ galaxy assigned based on the value of its physical property. The definition $WW(\theta)/DD(\theta)$ allows the estimation of MCF without requiring a random catalogue as it incorporates edge effects, eliminating concerns about the survey's geometry \citep{sheth_connolly_2005MCF, skibba2006MCF}. The estimator effectively calculates the correlation between galaxy properties and their environments, and it is expected to converge towards unity on large scales where the environmental influence on galaxy properties diminishes \citep{sheth2006environment, white2009breaking, skibba2012galaxy}.

\subsection{Assigning marks}
To compute MCFs, marks or weights must be assigned to galaxies based on the values of galaxy properties. Earlier studies used the absolute value of the physical property of interest as the mark \citep{sheth_connolly_2005MCF, skibba2006MCF}. However, this approach causes the MCF to depend on the distribution of the physical property, making it difficult to compare MCFs marked with different properties. This issue can be resolved by rank ordering the physical property and using the rank itself as the mark \citep{skibba2013measures}. The rank of a galaxy in a sample of $n$ galaxies will be a unique integer between $1$ and $n$, with ranks $1$ and $n$ corresponding to the galaxies with the lowest and highest values for the physical property, respectively. We then linearly rescale these ranks to a range between $0$ and $2$, so that the galaxy with the lowest value for the property (e.g., faintest galaxy) has rank $0$, and the galaxy with the highest value (e.g., brightest galaxy) has a rank of $2$.

The mark emphasizes galaxy pairs where both galaxies have higher property values. Thus, the MCF measured using Eq.~\ref{eq_mcf} shows whether the property’s spatial correlation is above or below the average, which is unity. A deviation from unity, referred to as the marked clustering signal or MCF signal, indicates how the number of such pairs deviates from what is expected by the standard two-point correlation function. A deviation greater than unity signifies an excess of pairs, indicating a positive correlation of the property, while a deviation less than unity suggests a reduced likelihood, reflecting an anti-correlation between the marking properties of galaxy pairs.

\subsection{Statistical Uncertainties}
The statistical errors of the angular two-point correlation function are estimated using the Jackknife resampling method \citep{norberg2009statistical} with Jackknife regions of size $\sim 5$ square degrees. The Jackknife covariance matrix is then 
\begin{equation}
          C_{ij} =\frac{ N-1}{N}\sum_{k=1}^{N} (\omega^k (\theta_i)-\bar\omega (\theta_i))(\omega^k (\theta_j)-\bar\omega (\theta_j))
         \end{equation}
where $\omega^k (\theta_i)$ is the angular correlation function at $\theta$=$\theta_i$ from the $k^\text{th}$ Jackknife region and $\bar\omega$ is the average correlation function of all the $N$ Jackknife regions \citep{Zehavi2005,scranton2002analysis}.

Simply summing the jackknife errors in $W(\theta)$ and $\omega(\theta)$ would overestimate the true error in the MCF because the errors in $W(\theta)$ are strongly correlated with those in $\omega(\theta)$. Therefore, to estimate the statistical errors in the MCF, we use the random-scrambling method \citep{skibba2006MCF, sheth_connolly_2005MCF}. Here, the error is estimated by remeasuring the MCF about 100 times after randomly scrambling the marks. The standard deviation around the mean of these measurements (which is unity) gives the uncertainty in the MCF.


\begin{figure*}
    \centering
    \includegraphics[width=1\linewidth]{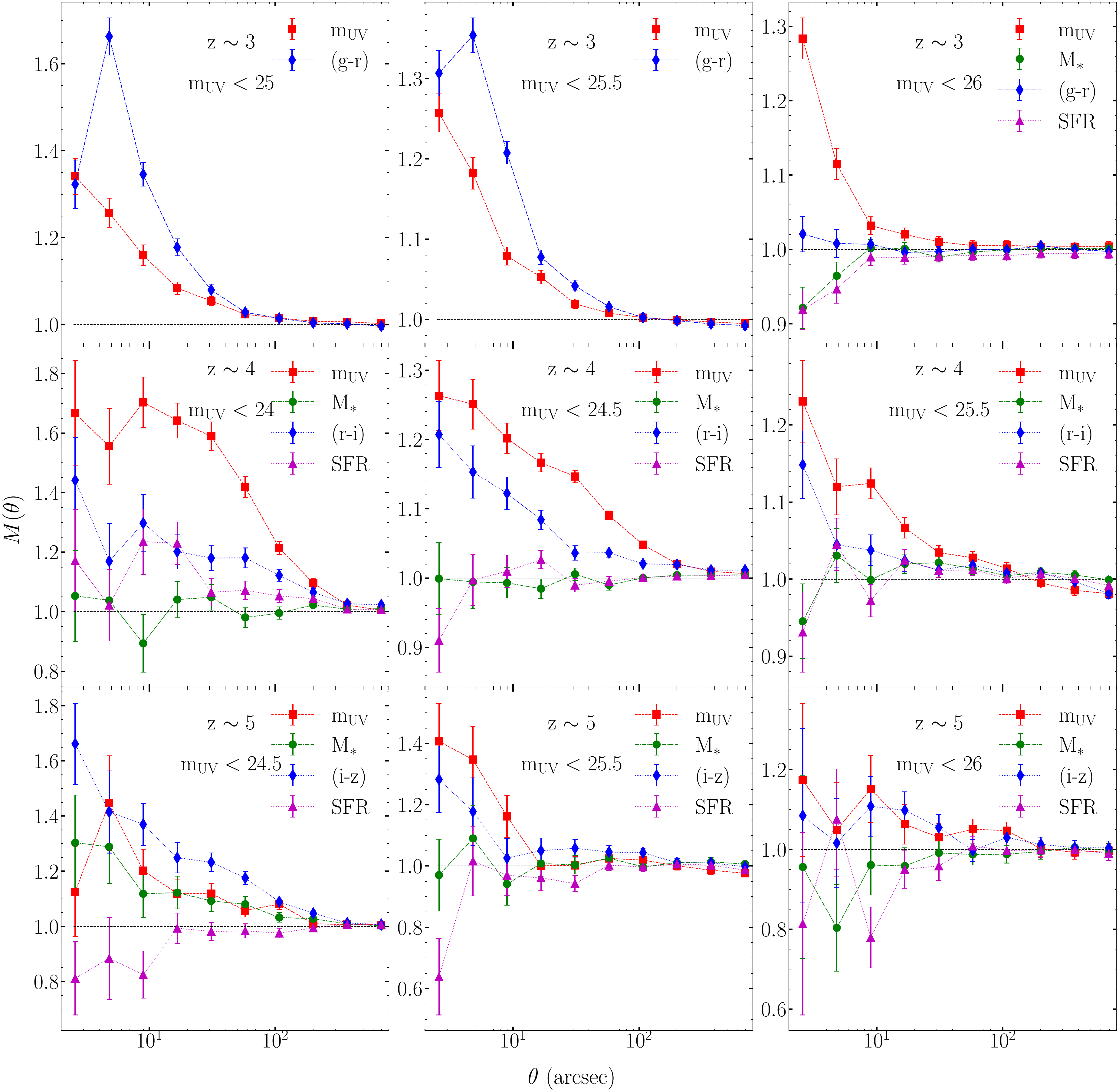}
    \caption{ The MCF of LBGs obtained by rank-ordering galaxies using different properties at $z\sim 3$, $z\sim 4$, and $z\sim 5$  for various threshold apparent magnitudes.}
    \label{fig_mcf}
\end{figure*}

\section{Results and Discussion} \label{sec_result}

In this section, we present the measurements of rank-ordered MCFs at $z \sim 3, 4,$ and $5$ for galaxy samples selected based on a threshold UV apparent magnitude $\text m_\text{UV}$. First, we investigate how different properties trace the galaxy environment at each redshift as a function of the separation between pairs. To compare the environmental dependence of galaxy properties across redshifts, we also present the MCF for samples with the same absolute magnitude, large-scale bias, and the average halo mass (inferred from the large-scale bias) at different $z$. The two-point angular correlation function of all the samples estimated using Eq.~\ref{eq_wtheta} is consistent with the measurements of \citet{harikane2022goldrush} using HSC-SSP public data release 2. %

\subsection{How LBG properties trace the environment?}
In Fig.~\ref{fig_mcf}, we present the MCF of LBGs derived by using the UV magnitude/luminosity, UV color, $M_\ast$, and SFR as marking properties. Since galaxy catalogs from the deep survey at $z \sim 3$ (corresponding to samples with threshold magnitude $m_{\text{UV}} < 25.5$) do not provide derived $M_\ast$ or SFR values, the marked correlation functions for these samples will not be presented using these properties as markers. For the ultra-deep catalogs, the derived $M_\ast$ and SFR  are provided at $z \sim 3$ ($m_{\text{UV}} < 26$), and the corresponding MCF measurements are shown in the top-right panel of Fig.~\ref{fig_mcf}. We first observe that for each redshift and threshold magnitude, the MCF amplitude varies significantly with the galaxy property used, clearly indicating that different galaxy properties trace the environment in distinct ways. Further, at all redshifts and for all galaxy properties, the deviation of the MCF from one is larger for brighter samples on all scales;  thus, the properties of brighter galaxies are generally more correlated with the environment than fainter ones.

It is also clear from the figure that at most redshifts and threshold magnitudes, UV magnitude-derived MCF differs substantially from unity. Notably at $z \sim 4$, for the threshold magnitudes $25.5$ and $26$ at $z \sim 5$, and for the faintest magnitude at $z \sim 3$, this measure shows the strongest signal compared to those based on other properties. This highlights UV luminosity/magnitude as one of the most effective tracers of the environment for high-z galaxies.

The MCF where galaxies are rank-ordered using UV colors (g-r at $z \sim 3$, r-i at $z \sim 4$, and i-z at $z \sim 5$) are also significant and comparable to those obtained from UV magnitude for many samples. 
Thus, the UV color is also a powerful tracer of the LBG environment. Since the UV spectral slope ($\beta$), defined by the relation $f_\lambda = f_0 \lambda^\beta$, is independent of UV magnitude \citep{Morales+2024_UVbeta, Finkelstein+2012_UVbeta}, the UV color exhibits a strong linear correlation with magnitude. Therefore, it is expected that the MCF using the UV color would follow that using the UV magnitude. However, the MCFs derived from UV magnitude and color do not precisely align. For brighter samples ($\text m_{UV} < 25.5$) at $z \sim 3$, the amplitude of MCFs using UV color is higher than that using magnitude, whereas the trend reverses at $z \sim 4$. The underlying cause of this difference remains unclear; however, the scatter in the color-magnitude relation may be a contributing factor.

On the other hand, the MCFs for magnitude-selected LBG samples, when galaxies are rank-ordered by their SFR and $M_\ast$, do not show significant deviations from unity compared to MCFs based on other observables. Therefore, these properties are less reliable tracers of the LBG environment. For the faintest sample at $z \sim 3$ and the brighter samples at $z \sim 4$ and $5$, the MCFs weighed using SFR or $M_\ast$ show a marginal departure from unity. Notably, the deviation at $z \sim 3$ is less than one when  SFR is the marking property whereas, at  $z \sim 4$ and $z \sim 5$, the deviations exceed one when SFR and $M_\ast$ are used as marking properties, respectively. 
This suggests that the LBGs differ from $z \sim 0$ galaxies, where SFR and  M$_\ast$ strongly trace the galaxy environment \citep{sureshkumar2021_gamaMCF}. Specifically, at $z \sim 0$, the  $M_\ast$ is the most reliable tracer of the environment, which contrasts with what we find at high-z.

The UV magnitude, originating from young stellar populations, serves as a reliable proxy for the SFR in galaxies with a constant star formation history \citep{Kennicutt1988_review, Madau+1998_sfh}. In this context, the SFR, like UV magnitude, should also be a robust indicator of the galactic environment. However, a more complicated star formation  (including rising, decaying, constant, and bursty phases) and the metal enrichment history for a galaxy can introduce a scatter in the UV magnitude - SFR relation \citep{Wilkins+2012_UV_SFR_relation, Dominguez+2015_burstySFH_SED}. This may eliminate the correlation between SFR and environment that was observed between UV magnitude and environment. 

Previous high-redshift studies by \citet{darvish+2016_sfr_quench_z1} and \citet{chartab+2020_SFR_env_candels} found that SFR decreases with increasing galaxy number density in the environment. In contrast, more recent studies \citep{shi+2024_sfr_env_z4, Taamoli+2023} observed a positive correlation between SFR and environmental density. These studies are based on the COSMOS2020 catalog \citep{weaver+2022_cosmos}, where $m_{\text{UV}}$ reaches a depth of $\sim 27.5$. The galaxy samples analyzed in this paper, derived from wide and deep surveys, are significantly brighter in UV magnitude compared to those used in these studies. Furthermore, while their samples are complete in either stellar mass or $K_s$-band magnitude, they are incomplete in UV magnitudes. 
Therefore, a comparison with these studies, if feasible, is restricted to the faintest samples (with magnitude threshold $\text m_\text{UV} < 26$) at $z \sim 3$ in our work. For this sample, we find the MCF of SFR to be below unity, indicating that galaxy pairs with high SFR are less common than ordinary pairs.  Nonetheless, a direct comparison remains challenging due to differences in sample selection criteria and statistic used, which may limit the conclusions that can be drawn. 

\begin{figure*}
    \centering
    \includegraphics[width=1\linewidth]{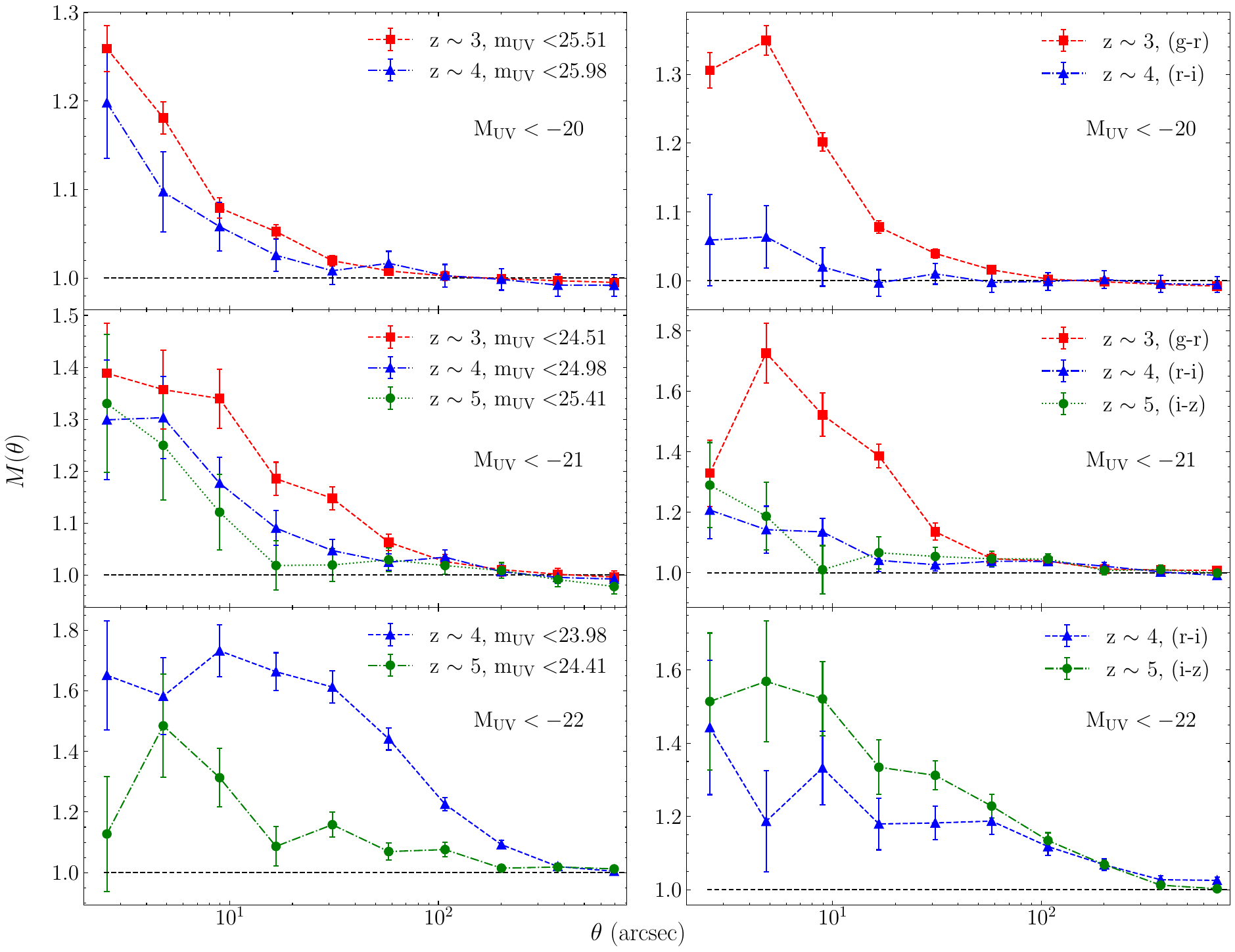}
    \caption{The MCF for galaxies at $z \sim 3$, $z \sim 4$, and $z \sim 5$ with the same threshold absolute magnitude across different redshifts. The left and right panels present the results for galaxies rank-ordered by UV luminosity and UV color, respectively.} 
    \label{fig_mcf_same_Muv}
\end{figure*}

\subsection{Dependence of MCF on Scale}
Fig.~\ref{fig_mcf} clearly shows that at all redshifts, the MCF measured using UV magnitude and color deviates from unity, even on angular scales extending up to $400''$. The marked clustering signal is most pronounced on small scales with $\theta \lesssim 10''$, where clustering within a single dark matter halo (the 1-halo term in the halo model) dominates \citep{jose+2013_lbg_acf, jose+2017_lbg_acf, harikane2022goldrush}. This strong signal on small scales is likely due to the correlation between properties of central-satellite and satellite-satellite galaxy pairs within a dark matter halo. For the brightest samples at every redshift, the marked clustering signal extends well beyond scales of $100''$, reaching up to $400''$ at $z \sim 4$. This scale is an order of magnitude larger than the size of a typical dark matter halo at these redshifts. For comparison, the virial radius of a large dark matter halo with a mass of $10^{12} \, M_\odot$  at $z\sim 4$, collapsing from $3 \sigma$ fluctuations, is approximately $300$ kpc, which corresponds to an angular scale of $40''$.

The MCF of galaxies at $z \sim 0$, measured using galaxy samples from the SDSS and GAMA surveys with $r-$band luminosities, also exhibits a deviation from unity on large scales \citep{skibba2013measures, sureshkumar2021_gamaMCF}. Notably, the deviation observed for high-z LBGs in this work is generally more pronounced than that at $z \sim 0$, both on small and large scales. This indicates that at high redshifts, there is an excess probability for galaxies with high luminosity or color to form pairs compared to those at low redshifts, which cannot be accounted for by the standard two-point correlation function.

The marked correlation function is often used to probe 2-halo galaxy conformity, which is the correlation between properties of central galaxies in distinct dark matter halos on scales bigger than a typical halo. \citet {CalderonV_BerlindA+2018} identified a weak yet statistically significant 2-halo conformity signal in low-z SDSS samples. While we find much stronger MCF signals on scales where 2-halo clustering is important, our measurements are not done on a sample of central galaxies, unlike those of \citet{CalderonV_BerlindA+2018}. Low-$z$ studies have pointed out that the physical origin of such large-scale signals could be attributed to factors such as the correlation between the halo mass function and environmental density \citep{skibba2006MCF}, halo/galaxy assembly bias \citep{paranjape+2015_conformity_conc, zu_2018_conformity_mcf}, etc. Further research is necessary to understand these more comprehensively.

\begin{figure*}
    \centering
    \includegraphics[width=0.8\linewidth]{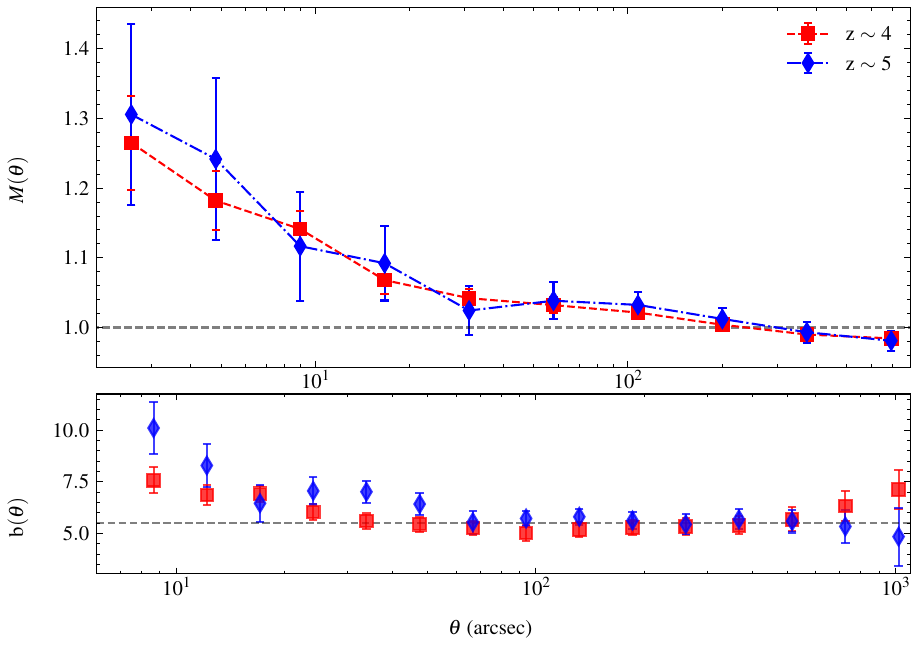}
    \caption{Top panel: UV magnitude rank-ordered MCF for galaxy samples with effective galaxy bias, $b_\text{eff} \sim 5.5$ at $z\sim 4$ and $5$. Bottom panel:  The galaxy bias of samples as a function of angular separation. The dashed horizontal line is the effective galaxy bias.} 
    \label{fig_mcf_same_bias}
\end{figure*}

\subsection{Samples with the same luminosity at different redshifts}

In this section, we examine how the environmental dependence of galaxy properties varies with redshift, for samples with the same threshold absolute magnitude ($\text M_{UV}$). Figure~\ref{fig_mcf_same_Muv} shows the UV magnitude and UV color marked MCFs for these galaxy samples at different redshifts. Note that MCF measurements are not shown for $z\sim  5$ with $M_{UV} = -20$ and $z\sim 3$ with $M_{UV} = -22$ due to large uncertainties. We find that, in general, the amplitude of the  MCFs using the magnitude increases from $z \sim 5$ to $\sim 3$ for all samples. Moreover, a comparison of results for samples with $M_{UV} > -20$ and $M_{UV} > -21$ reveals that the strength of the marked signal at $z \sim 3$, relative to that at $z\sim 4$ increases with the threshold magnitude of the sample. Similarly, the ratio of the MCF between $z \sim 4$ and $5$ is larger for the brightest sample.

When galaxies are rank-ordered by color, the environmental correlations of galaxy properties are found to be strongest at $z \sim 3$. For the fainter sample $( M_{\text{UV}} < -21 )$, the MCFs at $z \sim 4$ and $z \sim 5$ are comparable. On the other hand, for the brightest sample $(M_{\text{UV}} <-22 )$, the deviation of the MCF from unity is larger at $z \sim 5$ than at $z \sim 4$, differing from the behavior of MCFs based on magnitude at similar redshifts.

The evolution of the marked clustering signal with redshift for samples with the same absolute magnitude is intriguing. It is important to note that, although the absolute magnitudes of these samples are similar across redshifts, their dark matter halo and galaxy properties can differ. For example, clustering analysis of \citet{jose+2013_lbg_acf} finds that the average halo mass of the galaxy samples for the same absolute magnitude at $z\sim 3$ is higher by a factor $\sim 3$ compared to samples at $z \sim 5$.  Furthermore, the average $M_\ast$ of a high-z galaxy at a given UV luminosity evolves in the redshift range $3$ to $5$. At $M_{UV} = -20$ and $21$, the $M_\ast$ of the galaxies at $z\sim 3$ are larger by factors $1.5$ and $1.25$ respectively compared to that at $z\sim 5$ \citep{shibuya+2015_muv_mstar,  song+2016_Muv_Mstar, behroozi+2019_unimachine}.  Moreover, the average galaxy bias of these samples also evolves over time. The evolution of marked clustering signal with redshifts thus may be attributed to variations in dark matter halo characteristics and galaxy properties.

There are low redshift studies that measure the MCFs of samples complete in $M_\ast$ \citep{sureshkumar2021_gamaMCF}. This helps to gauge the importance of $M_\ast$ on the environmental dependence of galaxy properties. However, the high redshift samples used in this study are not $M_\ast$ complete. This is because a population of high star-forming, high $M_\ast$, and dusty galaxies at high-z that is not visible in UV is not included in our work,  as one requires FIR - submillimeter observations to probe these galaxies \citep{Harikane+2019_dusty_gals, zhang+2022_submm-highz-gals}. Moreover, as discussed in Section \ref{ssec_caveats}, the SFR and $M_\ast$ estimates in our samples suffer from various uncertainties. Thus, rather than examining the evolution of the MCF across redshift within $M_\ast$ bins, we construct galaxy samples at different redshifts with the same effective large-scale galaxy bias and average halo mass and investigate how the environmental dependence of the properties of these samples evolves with redshift.

\subsection{Samples with same effective halo bias} \label{sec_same_bias}

\begin{figure*}
    \centering
    \includegraphics[width=0.8\linewidth]{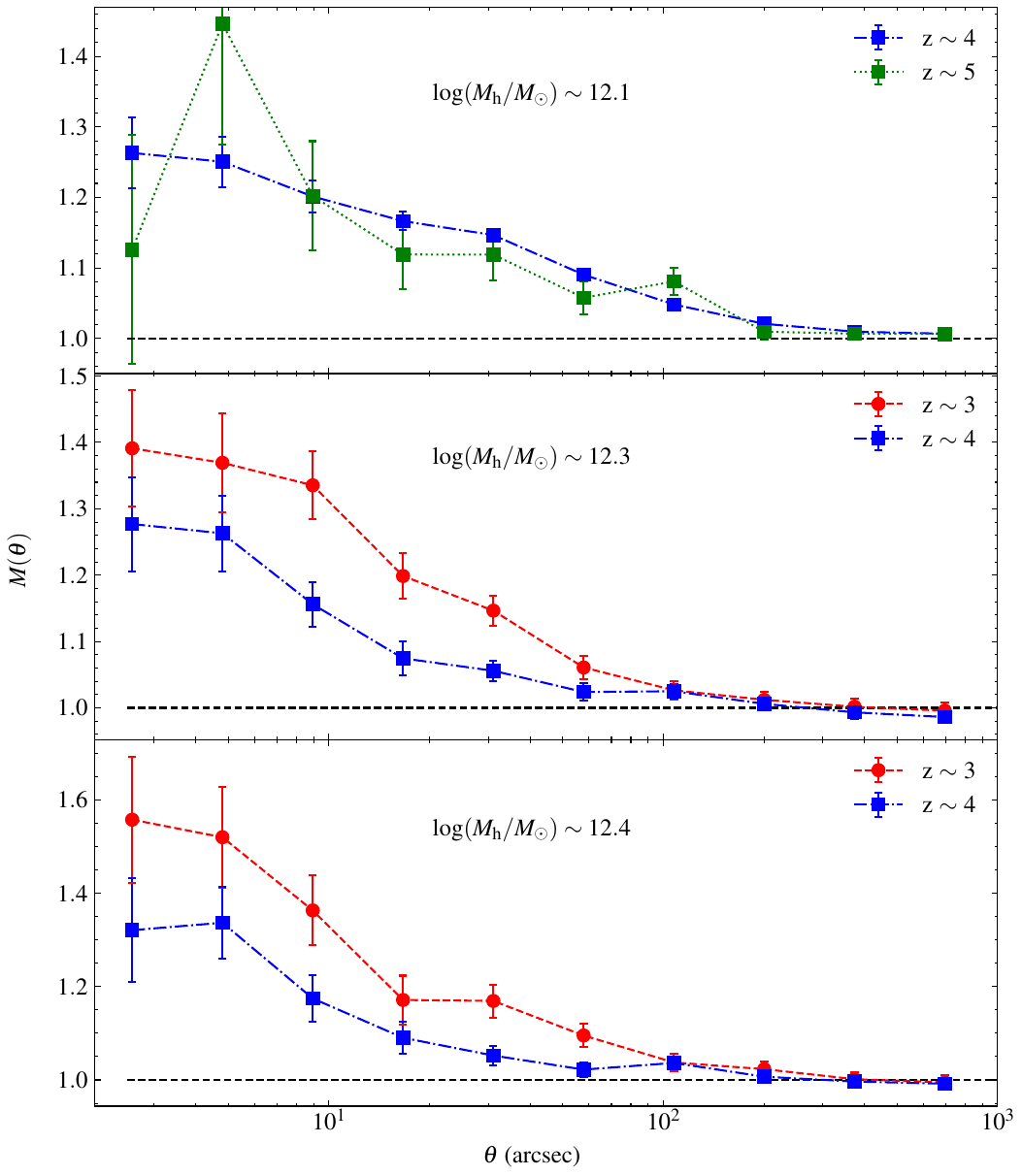}
     \caption{The UV magnitude rank-ordered MCFs of galaxy samples with the similar average halo mass at different redshifts. The top panel corresponds to wide survey whereas middle and bottom panel corresponds to deep survey.}
    \label{fig_mcf_same_mass}
\end{figure*}

In this section, we compare the magnitude-marked MCF of galaxy samples with the same effective large-scale galaxy bias at different redshifts. The large-scale bias, the ratio of the galaxy to dark matter correlation functions on large scales, measures the clustering of galaxies relative to that of matter. In linear theory, the large-scale bias is found to be a universal function of $\sigma(M,z)$ (expressed in terms of $\nu(M,z) = \delta_c/\sigma(M,z)$ where $\delta_c=1.686$), the rms variation of density contrast in comoving spheres of mass $M$ at $z$. Comparing the MCF of samples with the same large-scale bias allows us to assess whether the evolution of the MCF with redshift, especially on large scales, is linked to the sample's $\sigma(M,z)$. 

To compute an effective large-scale galaxy bias of the sample, we first estimate the galaxy angular correlation function as discussed in section~\ref{sec_mcf}. We also require the linear dark matter spatial correlation function given by 
\begin{equation}
    \xi_\text{mm}(r,z) = \dfrac{1}{2\pi^2}\int dk P(k,z) \dfrac{\sin(kr)}{kr}, 
\end{equation}
where $P(k,z)$ is the linear matter power spectrum, evaluated using  Planck cosmology \citep{Planck_2018}. 

By applying Limber transformation to $\xi_{\text{mm}}(r)$ and using the redshift distribution of sample galaxies given in Section~\ref{sec_data} \citep{Limber1953}, we estimate the dark matter angular correlation function $\omega_{\text{mm}}(\theta)$. The galaxy bias as a function of angular separation $\theta$, $b(\theta) = \omega(\theta)/\omega_{\text{mm}}(\theta)$, is found to be a constant over $100'' \leq  \theta \leq  500''$ for all galaxy samples. We define the effective large-scale galaxy bias as 
\begin{equation}
    b_\text{eff} = \dfrac{\int \omega(\theta) \, \theta^2 \, d\theta } {\int \omega_{\text{mm}}(\theta) \, \theta^2 \, d\theta}, 
    \
\end{equation}
where the integration ranges from $\theta = 100''$ to $500''$. 

In Fig.~\ref{fig_mcf_same_bias}, we compare the MCFs derived from the UV magnitude of samples from the deep region with approximately the same effective galaxy bias of $5.5$ at $z \sim 4$ and $5$. The apparent magnitudes of the samples with $b_\text{eff} \sim 5.5$ are $25.32$ and $25.35$ at $z \sim 4$ and $5$, respectively. In the lower panel, we also show the galaxy bias as a function of scale for these samples. We also measured MCFs for galaxy samples from the wide survey with the same effective bias; however, the signal strength was not statistically significant at every redshift, which limited the comparisons of these measurements between redshifts. 

Fig.~\ref{fig_mcf_same_bias} demonstrates that the MCFs of samples with $b_\text{eff} \sim 5.5$ at $z \sim 4$ and $z \sim 5$ are comparable. It is likely that the strength of the small-scale ($\theta \lesssim 10''$) MCF is heavily influenced by the processes of galaxy formation within a dark matter halo. Therefore, we compare the MCFs of galaxy bias-selected samples only on large scales. The similar large-scale MCF signals at $z \sim 4$ and $z \sim 5$ suggest a potential dependence on the large-scale galaxy bias; however, accurate measurements using multiple samples across different redshifts are needed to confirm this. 

\subsection{Samples with same effective halo mass} \label{sec_same_mass}

Galaxy samples at different redshifts with the same galaxy bias can possess varying halo masses since the halo mass-bias relationship evolves with redshift ($b = b(\sigma(M,z) )= b(M,z)$). Specifically, the large-scale bias of a galaxy sample of a given average halo mass increases with increasing redshift. Consequently, comparing the MCF of galaxy samples with similar average halo masses at different redshifts helps to determine the role of halo mass in the evolution of the MCF. 

Determining the average mass of a galaxy sample requires a comprehensive halo occupation distribution (HOD) analysis (see, for example, \citet{harikane2022goldrush}). However, at high redshifts, the halo mass derived from such analyses is constrained by uncertainties in the HOD model and a lack of precise understanding of halo clustering on quasi-linear scales. Therefore, we compare galaxy samples with the same `effective' halo mass across different redshifts. The effective halo mass of the sample, $M_\text{h}$, is determined from the large-scale bias (estimated in section~\ref{sec_same_bias}) using the halo mass-bias relation of \citet{Tinker+2010_bias}. Although this effective halo mass is not an exact estimate of the sample's average halo mass, it serves as a reliable indicator of the same.

By comparing the effective/average halo mass, estimated for samples with different threshold magnitudes, we obtained three galaxy samples with similar masses across different redshifts.  The masses of these samples are $\log(M_h/M_\odot) \sim 12.1$ at $z \sim 4$ and $5$, $\log(M_h/M_\odot) \sim 12.3$ at $z \sim 3$ and $4$, and $\log(M_h/M_\odot) \sim 12.4$ at $z \sim 3$ and $4$. The  magnitude-weighted MCFs of these samples are presented in Fig.~\ref{fig_mcf_same_mass}. The apparent magnitude threshold for the sample with $\log(M_h/M_\odot) = 12.1$ is $24.5$ at both $z \sim 4$ and $5$. For samples with $\log(M_h/M_\odot) = 12.3$ and $12.4$, the thresholds are $24.5$ and $24.3$ at $z \sim 3$, and $25.2$ and $25$ at $z \sim 4$, respectively. 

Figure~\ref{fig_mcf_same_mass} shows that for the galaxy sample with $\log(M_h/M_\odot) \sim 12.1$, the MCF signals at $z \sim 4$ has slightly higher amplitude than that at  $5$  for $\theta > 10''$. On smaller scales, measurement uncertainties prevent meaningful comparisons. For more massive samples, the environmental correlations are stronger at $z \sim 3$ than $4$. We do not show MCFs at $z \sim 5$ for these samples due to large of statistical uncertainties in measurements.  Overall, for samples selected by effective halo mass, the marked clustering strength increases as redshift decreases. 

\subsection{Systematics and Caveats} \label{ssec_caveats}
\begin{figure*}
\includegraphics[width=1\textwidth]{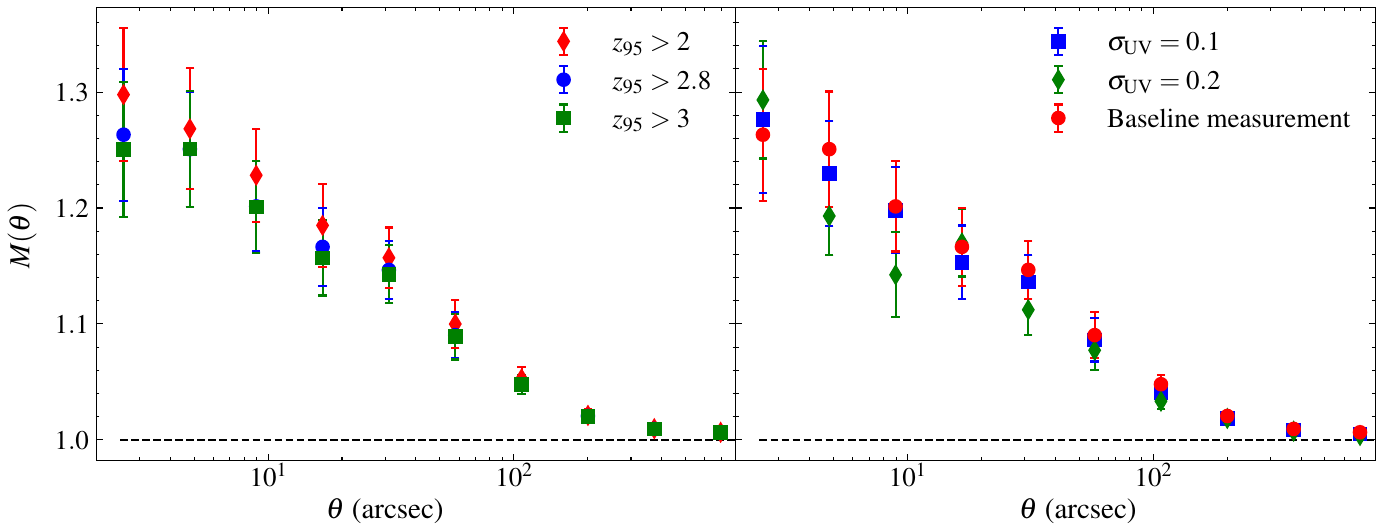}
\caption{Left panel: The MCFs of a wide-area sample at $z \sim 4$ for a magnitude threshold of $24.5$, analyzed with different criteria for removing low-$z$ interlopers. Right panel: The MCF for the same sample where a zero-mean Gaussian random noise with different standard deviation is added to the magnitudes.}\label{interlopers} 
\end{figure*}

The intrinsic UV luminosity of galaxies is commonly used as a proxy for SFR \citep{Kennicutt1988_review}. However, the SFR to UV luminosity calibration is influenced by assumptions on metal enrichment, dust attenuation, and the initial mass function (IMF) \citep{Mitchell+2013_Mstar_sfh_SED}. Variations in star formation history, whether bursty, rising, or declining, can cause significant changes (up to a factor of $\sim$ 2) in UV luminosities, theoretically derived from the SFR \citep{Dominguez+2015_burstySFH_SED, FVelazquez2021+_burstSFH_SED} \citep[see also][]{Kennicutt1988_review, Wilkins+2012_UV_SFR_relation, Cassara+2016_sfr_SED, Schaerer+2013_LBG_sfh_SED, Lee+2010_SFH_uncertainty}. 
%
%
Conversely, the SFR or M$_*$ inferred from broadband spectral energy distributions (SEDs) \citep{BejnaminJD+2013_SFR_estimate} are subject to biases and scatter. The MCF remains unaffected by systematic biases in derived SFR or $M_\ast$ that are independent of luminosity, as such biases do not alter the rank-ordering of galaxies. However, the scatter in the derived SFR or M$_*$ results in the weakening of the corresponding marked clustering signal. The photo-z derived using SED fitting are also susceptible to uncertainties \citep{tanaka2018photometric,nishizawa2020photometric}. 

\begin{figure}
\includegraphics[width=1\linewidth]{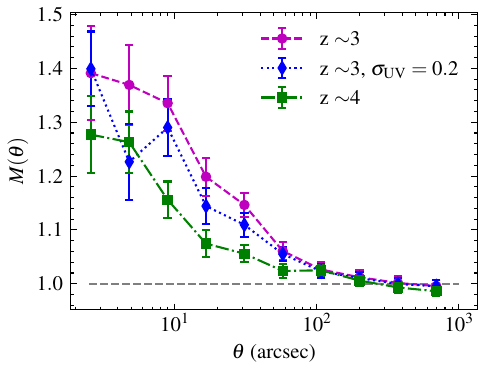}
\caption{ The MCF for sample with  $\log(M_h/M_\odot) \sim 12.3$ at $z\sim 3$ where noise with $\sigma_\text{UV} = 0.2$ has been added to the magnitudes together with MCFs at $z \sim 3$ and $z \sim 4$ without any added noise (same as the MCFs given in the middle panel of Fig.~\ref{fig_mcf_same_mass}).}\label{fig_mcf_err_mag_mass_samples} 
\end{figure}

The presence of low-$z$ interlopers with unknown clustering strength could introduce potential biases in the measurements, an effect that is neglected in the angular correlation function analysis of \citep{harikane2022goldrush}. To evaluate the impact of low-$z$ interlopers, we analyzed the magnitude-weighted MCFs for three wide-area g-dropout samples with a magnitude threshold of $24.5$, defined by photometric redshift constraints of $z_{95} > 2$, $z_{95} > 2.8$, and $z_{95} > 3$, where $z_{95}$ represents the upper bound of the 95\% confidence interval for the photometric redshift. The results shown in the left panel of Fig.~\ref{interlopers}, indicate that the MCFs do not change significantly between the samples, suggesting that the measured environmental correlations are largely robust against variations in the low-z interloper selection criteria. 

We also investigated the effect of uncertainties in observed magnitudes on the measured MCFs by adding Gaussian random noise with a mean of $0$ and standard deviations of $\sigma_\text{UV} = 0.1$ and $0.2$ to the galaxy magnitudes. After adding the noise, the galaxies were re-ranked based on the modified magnitudes, and the MCF was recalculated. The right panel of Figure~\ref{interlopers} shows the resulting MCFs at $z \sim 4$ after adding noise, alongside the baseline MCF for the sample with a threshold magnitude of $24.5$, where no uncertainties were added to the magnitudes. As expected, the MCF of the samples with added uncertainties in the observed magnitudes is lower than the baseline  measurements. However, a strong signal persists even for $\sigma_\text{UV} = 0.2$, demonstrating the robustness of MCF measurements to such uncertainties. 

We also note that the differences in MCFs at different redshifts (as presented in Figs.~\ref{fig_mcf_same_Muv} and \ref{fig_mcf_same_mass}) could partly be attributed to the varying uncertainties in the measured magnitudes across these redshifts. To illustrate this, in Fig.~\ref{fig_mcf_err_mag_mass_samples}, we show the MCF for $z \sim 3$ (dotted blue), where noise with $\sigma_\text{UV} = 0.2$ has been added to the magnitudes. Also shown in the figure are the noise-free MCFs at both $z \sim 3$ and $z \sim 4$ (as shown in the middle panel of Fig.~\ref{fig_mcf_same_mass}) for samples with $M_\text{h}/M_\odot \sim 12.3$. It is clear that the difference between the MCFs at $z \sim 4$ and $z \sim 3$, with added noise, is smaller.

\section{Summary and Conclusions} \label{sec_summary}

In this paper, we use rank-ordered MCF statistic to observationally probe how the properties of LBGs in the redshift range $z \sim 3$ to $5$ correlate with their environment. Our analysis utilizes data from the ultra-deep, deep and wide surveys of the CLAUDS and HSC-SSP programs( See table~\ref{table1}) . Taking advantage of a survey area covering approximately 600 square degrees at these redshifts, we are able to examine the environmental correlations of brighter LBG samples compared to previous studies that used alternative statistics. LBG samples are selected using a UV magnitude threshold, and MCFs are calculated by rank-ordering galaxies according to their UV luminosity, UV color, $M_\ast$, and SFR. The key findings from the measured MCFs are as follows: 

\begin{itemize}
    \item Different LBG properties trace the LBG environment in distinct ways. The UV magnitude is a strong tracer of the LBG environment. The MCF obtained using UV colors (g-r at $z \sim 3$, r-i at $z \sim 4$, and i-z at $z \sim 5$) as markers are strong and comparable to those derived using UV luminosity, though they do not precisely align. MCFs using SFR and $M_\ast$ show less deviation from unity, making them less reliable tracers of LBG environments. 
    \item The UV magnitude and color-derived MCFs for brighter LBGs exhibit larger deviations from unity across all redshifts, indicating a strong environmental correlation of these properties in brighter samples.
  
     \item  The marked signal is most pronounced on small scales ($\theta \leq 10''$) corresponding to the size of typical dark matter halos, and hence, the signal is likely due to interactions between central and satellite galaxies in a dark matter halo. The MCF deviates from unity even on larger scales ($10 \leq \theta \leq 100''$), and the signal persists even up to $400''$ at $z \sim 3$ and $4$. The presence of a large-scale signal could be evidence of 2-halo conformity, though further research is needed to fully understand these signals and their origins. 
     \item The MCFs of LBGs using the UV magnitude and color show a stronger deviation from unity compared to the $r-$ band luminosity derived MCFs of $z \sim 0 $ galaxy samples from SDSS and GAMA \citep{skibba2013measures, sureshkumar2021_gamaMCF}. 
\end{itemize}

To compare galaxy populations across different redshifts, we analyzed the MCFs of populations with identical absolute magnitudes and observed the following:

\begin{itemize}

\item The environmental correlations of galaxy magnitudes generally increases from $z \sim 5$ to $z \sim 3$ for all samples.

\item The MCF signal is strongest at $z \sim 3$ when galaxies are ranked by color. For the brightest sample ($M_\text{UV} = -22$), the deviation of the MCF from unity is more pronounced at $z \sim 5$ than at $z \sim 4$, which contrasts with the MCFs ranked by magnitude.

\end{itemize}

Despite similar absolute magnitudes, galaxy samples exhibit significant variations in dark matter halo and galaxy properties - such as large-scale bias and effective halo mass- across redshifts. These variations could explain the observed differences in marked clustering signals. To control for these effects, we compared the MCFs of samples at different redshifts with comparable effective large-scale galaxy bias and average halo mass (defined in Sections \ref{sec_same_bias} and \ref{sec_same_mass}), leading to the following findings.

\begin{itemize}
    
   \item For samples with effective galaxy bias $5.5$ at $z \sim 3$ and $4$, the MCF signals are comparable on large scales, suggesting a potential dependence of large-scale MCF on galaxy bias.
   \item For samples selected by effective halo mass, the marked clustering strength generally increases  as the redshift decreases from $5$ to $3$.
\end{itemize}

Our measurements present the first exploration of environmental correlations of galaxy properties within the redshift range of $3$ to $5$ using the MCF statistic. Additionally, we leverage brighter galaxy samples compared to previous studies that employed alternative statistical methods for probing environmental correlations. Comparing our key findings—including the strong environmental correlations of certain properties in high-redshift LBGs, the evolution of these correlations with redshift, and their potential link to galaxy bias and mass—with simulations and semi-analytic galaxy formation models is expected to provide tighter constraints on theoretical models (see, e.g., \citep{skibba2013measures, CalderonV_BerlindA+2018, SFisher+_2022, szewciw+2022} for MCF models using semi-analytic approaches and N-body simulations that probe the environmental correlations of galaxies at low $z$). We plan to explore these in future work.

\section*{Acknowledgements}
We are grateful to the anonymous referee for the detailed and thoughtful comments. The authors thank Yuichi Harikane for providing us with galaxy samples from PDR2 for analysis and clarifying doubts. CJ is supported by the University Grant Commission, India, through a BSR start-up grant (F.30-463/2019(BSR)) and the Rashtriya Uchchatar Shiksha Abhiyan scheme (RUSA 2.0, project number: T3A). EM acknowledges the financial support in the form of the fellowship of Cochin University of Science and Technology, Kerala and RUSA 2.0 (project number: T3A).
\bibliographystyle{mnras}
\bibliography{main}

\begin{thebibliography}{}
\makeatletter
\relax
\def\mn@urlcharsother{\let\do\@makeother \do\$\do\&\do\#\do\^\do\_\do\%\do\~}
\def\mn@doi{\begingroup\mn@urlcharsother \@ifnextchar [ {\mn@doi@}
  {\mn@doi@[]}}
\def\mn@doi@[#1]#2{\def\@tempa{#1}\ifx\@tempa\@empty \href
  {http://dx.doi.org/#2} {doi:#2}\else \href {http://dx.doi.org/#2} {#1}\fi
  \endgroup}
\def\mn@eprint#1#2{\mn@eprint@#1:#2::\@nil}
\def\mn@eprint@arXiv#1{\href {http://arxiv.org/abs/#1} {{\tt arXiv:#1}}}
\def\mn@eprint@dblp#1{\href {http://dblp.uni-trier.de/rec/bibtex/#1.xml}
  {dblp:#1}}
\def\mn@eprint@#1:#2:#3:#4\@nil{\def\@tempa {#1}\def\@tempb {#2}\def\@tempc
  {#3}\ifx \@tempc \@empty \let \@tempc \@tempb \let \@tempb \@tempa \fi \ifx
  \@tempb \@empty \def\@tempb {arXiv}\fi \@ifundefined
  {mn@eprint@\@tempb}{\@tempb:\@tempc}{\expandafter \expandafter \csname
  mn@eprint@\@tempb\endcsname \expandafter{\@tempc}}}

\bibitem[\protect\citeauthoryear{{Aihara} et~al.,}{{Aihara}
  et~al.}{2018}]{aihara+2018_pdr1_subaru}
{Aihara} H.,  et~al., 2018, \mn@doi [\pasj] {10.1093/pasj/psx081}, \href
  {https://ui.adsabs.harvard.edu/abs/2018PASJ...70S...8A} {70, S8}

\bibitem[\protect\citeauthoryear{Aihara et~al.,}{Aihara
  et~al.}{2019}]{aihara2019second}
Aihara H.,  et~al., 2019, Publications of the Astronomical Society of Japan,
  71, 114

\bibitem[\protect\citeauthoryear{{Aihara} et~al.,}{{Aihara}
  et~al.}{2022}]{aihara+2022_pdr3_subaru}
{Aihara} H.,  et~al., 2022, \mn@doi [\pasj] {10.1093/pasj/psab122}, \href
  {https://ui.adsabs.harvard.edu/abs/2022PASJ...74..247A} {74, 247}

\bibitem[\protect\citeauthoryear{{Armijo}, {Cai}, {Padilla}, {Li}  \&
  {Peacock}}{{Armijo} et~al.}{2018}]{Armijo2018}
{Armijo} J.,  {Cai} Y.-C.,  {Padilla} N.,  {Li} B.,   {Peacock} J.~A.,  2018,
  \mn@doi [\mnras] {10.1093/mnras/sty1335}, \href
  {https://ui.adsabs.harvard.edu/abs/2018MNRAS.478.3627A} {478, 3627}

\bibitem[\protect\citeauthoryear{Arnold}{Arnold}{1995}]{stoyan1994fractals}
Arnold L.,  1995, \mn@doi [ZAMM - Journal of Applied Mathematics and Mechanics
  / Zeitschrift für Angewandte Mathematik und Mechanik]
  {https://doi.org/10.1002/zamm.19950750815}, 75, 614

\bibitem[\protect\citeauthoryear{{Arnouts} \& {Ilbert}}{{Arnouts} \&
  {Ilbert}}{2011}]{Arnouts+2011_lephare}
{Arnouts} S.,  {Ilbert} O.,  2011, {LePHARE: Photometric Analysis for Redshift
  Estimate}, Astrophysics Source Code Library, record ascl:1108.009

\bibitem[\protect\citeauthoryear{{Aviles}}{{Aviles}}{2021}]{Aviles2021}
{Aviles} A.,  2021, \mn@doi [arXiv e-prints] {10.48550/arXiv.2110.13767}, \href
  {https://ui.adsabs.harvard.edu/abs/2021arXiv211013767A} {p. arXiv:2110.13767}

\bibitem[\protect\citeauthoryear{{Aviles}, {Koyama}, {Cervantes-Cota},
  {Winther}  \& {Li}}{{Aviles} et~al.}{2020}]{Aviles2020}
{Aviles} A.,  {Koyama} K.,  {Cervantes-Cota} J.~L.,  {Winther} H.~A.,   {Li}
  B.,  2020, \mn@doi [\jcap] {10.1088/1475-7516/2020/01/006}, \href
  {https://ui.adsabs.harvard.edu/abs/2020JCAP...01..006A} {2020, 006}

\bibitem[\protect\citeauthoryear{{Behroozi}, {Wechsler}, {Hearin}  \&
  {Conroy}}{{Behroozi} et~al.}{2019}]{behroozi+2019_unimachine}
{Behroozi} P.,  {Wechsler} R.~H.,  {Hearin} A.~P.,   {Conroy} C.,  2019,
  \mn@doi [\mnras] {10.1093/mnras/stz1182}, \href
  {https://ui.adsabs.harvard.edu/abs/2019MNRAS.488.3143B} {488, 3143}

\bibitem[\protect\citeauthoryear{{Beisbart} \& {Kerscher}}{{Beisbart} \&
  {Kerscher}}{2000}]{beisbart+2000}
{Beisbart} C.,  {Kerscher} M.,  2000, \mn@doi [\apj] {10.1086/317788}, \href
  {https://ui.adsabs.harvard.edu/abs/2000ApJ...545....6B} {545, 6}

\bibitem[\protect\citeauthoryear{Bertin \& Arnouts}{Bertin \&
  Arnouts}{1996}]{bertin1996sextractor}
Bertin E.,  Arnouts S.,  1996, Astronomy and astrophysics supplement series,
  117, 393

\bibitem[\protect\citeauthoryear{{Bosch} et~al.,}{{Bosch}
  et~al.}{2018}]{Bosch+2018_hscpipe}
{Bosch} J.,  et~al., 2018, \mn@doi [\pasj] {10.1093/pasj/psx080}, \href
  {https://ui.adsabs.harvard.edu/abs/2018PASJ...70S...5B} {70, S5}

\bibitem[\protect\citeauthoryear{{Calderon}, {Berlind}  \& {Sinha}}{{Calderon}
  et~al.}{2018}]{CalderonV_BerlindA+2018}
{Calderon} V.~F.,  {Berlind} A.~A.,   {Sinha} M.,  2018, \mn@doi [\mnras]
  {10.1093/mnras/sty2000}, \href
  {https://ui.adsabs.harvard.edu/abs/2018MNRAS.480.2031C} {480, 2031}

\bibitem[\protect\citeauthoryear{Cassarà et~al.,}{Cassarà
  et~al.}{2016}]{Cassara+2016_sfr_SED}
Cassarà L.~P.,  et~al., 2016, \mn@doi [Astronomy &amp; Astrophysics]
  {10.1051/0004-6361/201526505}, 593, A9

\bibitem[\protect\citeauthoryear{{Chartab} et~al.,}{{Chartab}
  et~al.}{2020}]{chartab+2020_SFR_env_candels}
{Chartab} N.,  et~al., 2020, \mn@doi [\apj] {10.3847/1538-4357/ab61fd}, \href
  {https://ui.adsabs.harvard.edu/abs/2020ApJ...890....7C} {890, 7}

\bibitem[\protect\citeauthoryear{{Cochrane}, {Best}, {Sobral}, {Smail},
  {Geach}, {Stott}  \& {Wake}}{{Cochrane}
  et~al.}{2018}]{Cochrane+2018_xir_depend}
{Cochrane} R.~K.,  {Best} P.~N.,  {Sobral} D.,  {Smail} I.,  {Geach} J.~E.,
  {Stott} J.~P.,   {Wake} D.~A.,  2018, \mn@doi [\mnras]
  {10.1093/mnras/stx3345}, \href
  {https://ui.adsabs.harvard.edu/abs/2018MNRAS.475.3730C} {475, 3730}

\bibitem[\protect\citeauthoryear{{Coil} et~al.,}{{Coil}
  et~al.}{2008}]{Coli_Newman+2008_color_xir_deep2}
{Coil} A.~L.,  et~al., 2008, \mn@doi [\apj] {10.1086/523639}, \href
  {https://ui.adsabs.harvard.edu/abs/2008ApJ...672..153C} {672, 153}

\bibitem[\protect\citeauthoryear{{Cooke}, {Omori}  \& {Ryan-Weber}}{{Cooke}
  et~al.}{2013}]{Cook+2013_LBG_env_2pcf}
{Cooke} J.,  {Omori} Y.,   {Ryan-Weber} E.,  2013, in American Astronomical
  Society Meeting Abstracts. p. 208.07

\bibitem[\protect\citeauthoryear{{Cooper} et~al.,}{{Cooper}
  et~al.}{2008}]{Cooper+2008}
{Cooper} M.~C.,  et~al., 2008, \mn@doi [\mnras]
  {10.1111/j.1365-2966.2007.12613.x}, \href
  {https://ui.adsabs.harvard.edu/abs/2008MNRAS.383.1058C} {383, 1058}

\bibitem[\protect\citeauthoryear{Coupon, Czakon, Bosch, Komiyama, Medezinski,
  Miyazaki  \& Oguri}{Coupon et~al.}{2018}]{coupon2018bright}
Coupon J.,  Czakon N.,  Bosch J.,  Komiyama Y.,  Medezinski E.,  Miyazaki S.,
  Oguri M.,  2018, Publications of the Astronomical Society of Japan, 70, S7

\bibitem[\protect\citeauthoryear{{Darvish}, {Mobasher}, {Sobral}, {Scoville}
  \& {Aragon-Calvo}}{{Darvish} et~al.}{2015}]{Darvish+2015_galaxy_env_z3}
{Darvish} B.,  {Mobasher} B.,  {Sobral} D.,  {Scoville} N.,   {Aragon-Calvo}
  M.,  2015, \mn@doi [\apj] {10.1088/0004-637X/805/2/121}, \href
  {https://ui.adsabs.harvard.edu/abs/2015ApJ...805..121D} {805, 121}

\bibitem[\protect\citeauthoryear{{Darvish}, {Mobasher}, {Sobral}, {Rettura},
  {Scoville}, {Faisst}  \& {Capak}}{{Darvish}
  et~al.}{2016}]{darvish+2016_sfr_quench_z1}
{Darvish} B.,  {Mobasher} B.,  {Sobral} D.,  {Rettura} A.,  {Scoville} N.,
  {Faisst} A.,   {Capak} P.,  2016, \mn@doi [\apj]
  {10.3847/0004-637X/825/2/113}, \href
  {https://ui.adsabs.harvard.edu/abs/2016ApJ...825..113D} {825, 113}

\bibitem[\protect\citeauthoryear{{Desprez} et~al.,}{{Desprez}
  et~al.}{2023}]{desprez+2023_clauds}
{Desprez} G.,  et~al., 2023, \mn@doi [\aap] {10.1051/0004-6361/202243363},
  \href {https://ui.adsabs.harvard.edu/abs/2023A&A...670A..82D} {670, A82}

\bibitem[\protect\citeauthoryear{Dom{\'\i}nguez, Siana, Brooks, Christensen,
  Bruzual, Stark  \& Alavi}{Dom{\'\i}nguez
  et~al.}{2015}]{Dominguez+2015_burstySFH_SED}
Dom{\'\i}nguez A.,  Siana B.,  Brooks A.~M.,  Christensen C.~R.,  Bruzual G.,
  Stark D.~P.,   Alavi A.,  2015, Mon. Not. R. Astron. Soc., 451, 839

\bibitem[\protect\citeauthoryear{{Durkalec} et~al.,}{{Durkalec}
  et~al.}{2018}]{Durkalec+2018_xir_depend_vimos}
{Durkalec} A.,  et~al., 2018, \mn@doi [\aap] {10.1051/0004-6361/201730734},
  \href {https://ui.adsabs.harvard.edu/abs/2018A&A...612A..42D} {612, A42}

\bibitem[\protect\citeauthoryear{{Euclid Collaboration} et~al.,}{{Euclid
  Collaboration} et~al.}{2020}]{desprez+2020_euclid}
{Euclid Collaboration} et~al., 2020, \mn@doi [\aap]
  {10.1051/0004-6361/202039403}, \href
  {https://ui.adsabs.harvard.edu/abs/2020A&A...644A..31E} {644, A31}

\bibitem[\protect\citeauthoryear{{Finkelstein} et~al.,}{{Finkelstein}
  et~al.}{2012}]{Finkelstein+2012_UVbeta}
{Finkelstein} S.~L.,  et~al., 2012, \mn@doi [\apj]
  {10.1088/0004-637X/756/2/164}, \href
  {https://ui.adsabs.harvard.edu/abs/2012ApJ...756..164F} {756, 164}

\bibitem[\protect\citeauthoryear{Flores~Vel{\'a}zquez
  et~al.,}{Flores~Vel{\'a}zquez et~al.}{2021}]{FVelazquez2021+_burstSFH_SED}
Flores~Vel{\'a}zquez J.~A.,  et~al., 2021, Mon. Not. R. Astron. Soc., 501, 4812

\bibitem[\protect\citeauthoryear{Furusawa et~al.,}{Furusawa
  et~al.}{2018}]{furusawa2018site}
Furusawa H.,  et~al., 2018, Publications of the Astronomical Society of Japan,
  70, S3

\bibitem[\protect\citeauthoryear{Giavalisco}{Giavalisco}{2002}]{giavalisco2002lyman}
Giavalisco M.,  2002, Annual Review of Astronomy and Astrophysics, 40, 579

\bibitem[\protect\citeauthoryear{{Grogin} et~al.,}{{Grogin}
  et~al.}{2011}]{Grogin+2011_candels}
{Grogin} N.~A.,  et~al., 2011, \mn@doi [\apjs] {10.1088/0067-0049/197/2/35},
  \href {https://ui.adsabs.harvard.edu/abs/2011ApJS..197...35G} {197, 35}

\bibitem[\protect\citeauthoryear{{Gu}, {Fang}, {Yuan}, {Lu}  \& {Liu}}{{Gu}
  et~al.}{2021}]{Gu+2021_env_SFR_candels}
{Gu} Y.,  {Fang} G.,  {Yuan} Q.,  {Lu} S.,   {Liu} S.,  2021, \mn@doi [\apj]
  {10.3847/1538-4357/ac1ce0}, \href
  {https://ui.adsabs.harvard.edu/abs/2021ApJ...921...60G} {921, 60}

\bibitem[\protect\citeauthoryear{{Guo} et~al.,}{{Guo}
  et~al.}{2013}]{Guo_zehavi+2013_xir_color_sdss}
{Guo} H.,  et~al., 2013, \mn@doi [\apj] {10.1088/0004-637X/767/2/122}, \href
  {https://ui.adsabs.harvard.edu/abs/2013ApJ...767..122G} {767, 122}

\bibitem[\protect\citeauthoryear{Harikane et~al.,}{Harikane
  et~al.}{2018}]{harikane2018goldrush}
Harikane Y.,  et~al., 2018, Publications of the Astronomical Society of Japan,
  70, S11

\bibitem[\protect\citeauthoryear{{Harikane} et~al.,}{{Harikane}
  et~al.}{2019}]{Harikane+2019_dusty_gals}
{Harikane} Y.,  et~al., 2019, \mn@doi [\apj] {10.3847/1538-4357/ab2cd5}, \href
  {https://ui.adsabs.harvard.edu/abs/2019ApJ...883..142H} {883, 142}

\bibitem[\protect\citeauthoryear{Harikane et~al.,}{Harikane
  et~al.}{2022}]{harikane2022goldrush}
Harikane Y.,  et~al., 2022, The Astrophysical Journal Supplement Series, 259,
  20

\bibitem[\protect\citeauthoryear{Harker, Cole, Helly, Frenk  \& Jenkins}{Harker
  et~al.}{2006}]{harker2006marked}
Harker G.,  Cole S.,  Helly J.,  Frenk C.,   Jenkins A.,  2006, Monthly Notices
  of the Royal Astronomical Society, 367, 1039

\bibitem[\protect\citeauthoryear{{Hartzenberg}, {Cowley}, {Hopkins}  \&
  {Allen}}{{Hartzenberg} et~al.}{2023}]{Hartzenberg+2023_sfr_quench_z2.5}
{Hartzenberg} G.~R.,  {Cowley} M.~J.,  {Hopkins} A.~M.,   {Allen} R.~J.,  2023,
  \mn@doi [\pasa] {10.1017/pasa.2023.42}, \href
  {https://ui.adsabs.harvard.edu/abs/2023PASA...40...43H} {40, e043}

\bibitem[\protect\citeauthoryear{{Hern{\'a}ndez-Aguayo}, {Baugh}  \&
  {Li}}{{Hern{\'a}ndez-Aguayo} et~al.}{2018}]{Hernandez2018}
{Hern{\'a}ndez-Aguayo} C.,  {Baugh} C.~M.,   {Li} B.,  2018, \mn@doi [\mnras]
  {10.1093/mnras/sty1822}, \href
  {https://ui.adsabs.harvard.edu/abs/2018MNRAS.479.4824H} {479, 4824}

\bibitem[\protect\citeauthoryear{Hsieh \& Yee}{Hsieh \&
  Yee}{2014a}]{hsieh2014estimating}
Hsieh B.,  Yee H.,  2014a, The Astrophysical Journal, 792, 102

\bibitem[\protect\citeauthoryear{{Hsieh} \& {Yee}}{{Hsieh} \&
  {Yee}}{2014b}]{hsieh+2014_demp}
{Hsieh} B.~C.,  {Yee} H.~K.~C.,  2014b, \mn@doi [\apj]
  {10.1088/0004-637X/792/2/102}, \href
  {https://ui.adsabs.harvard.edu/abs/2014ApJ...792..102H} {792, 102}

\bibitem[\protect\citeauthoryear{Huang et~al.,}{Huang
  et~al.}{2018}]{huang2018characterization}
Huang S.,  et~al., 2018, Publications of the Astronomical Society of Japan, 70,
  S6

\bibitem[\protect\citeauthoryear{Johnson et~al.,}{Johnson
  et~al.}{2013}]{BejnaminJD+2013_SFR_estimate}
Johnson B.~D.,  et~al., 2013, \mn@doi [The Astrophysical Journal]
  {10.1088/0004-637x/772/1/8}, 772, 8

\bibitem[\protect\citeauthoryear{{Jose}, {Subramanian}, {Srianand}  \&
  {Samui}}{{Jose} et~al.}{2013}]{jose+2013_lbg_acf}
{Jose} C.,  {Subramanian} K.,  {Srianand} R.,   {Samui} S.,  2013, \mn@doi
  [\mnras] {10.1093/mnras/sts503}, \href
  {https://ui.adsabs.harvard.edu/abs/2013MNRAS.429.2333J} {429, 2333}

\bibitem[\protect\citeauthoryear{{Jose}, {Baugh}, {Lacey}  \&
  {Subramanian}}{{Jose} et~al.}{2017}]{jose+2017_lbg_acf}
{Jose} C.,  {Baugh} C.~M.,  {Lacey} C.~G.,   {Subramanian} K.,  2017, \mn@doi
  [\mnras] {10.1093/mnras/stx1014}, \href
  {https://ui.adsabs.harvard.edu/abs/2017MNRAS.469.4428J} {469, 4428}

\bibitem[\protect\citeauthoryear{{Kauffmann}, {White}, {Heckman}, {M{\'e}nard},
  {Brinchmann}, {Charlot}, {Tremonti}  \& {Brinkmann}}{{Kauffmann}
  et~al.}{2004}]{Kauffmann+2004}
{Kauffmann} G.,  {White} S. D.~M.,  {Heckman} T.~M.,  {M{\'e}nard} B.,
  {Brinchmann} J.,  {Charlot} S.,  {Tremonti} C.,   {Brinkmann} J.,  2004,
  \mn@doi [\mnras] {10.1111/j.1365-2966.2004.08117.x}, \href
  {https://ui.adsabs.harvard.edu/abs/2004MNRAS.353..713K} {353, 713}

\bibitem[\protect\citeauthoryear{Kawanomoto et~al.,}{Kawanomoto
  et~al.}{2018}]{kawanomoto2018hyper}
Kawanomoto S.,  et~al., 2018, Publications of the Astronomical Society of
  Japan, 70, 66

\bibitem[\protect\citeauthoryear{{Kennicutt}}{{Kennicutt}}{1998}]{Kennicutt1988_review}
{Kennicutt} Robert~C. J.,  1998, \mn@doi [\araa]
  {10.1146/annurev.astro.36.1.189}, \href
  {https://ui.adsabs.harvard.edu/abs/1998ARA&A..36..189K} {36, 189}

\bibitem[\protect\citeauthoryear{{Koyama} et~al.,}{{Koyama}
  et~al.}{2018}]{Koyama+2018_sfr_env_subaru}
{Koyama} Y.,  et~al., 2018, \mn@doi [\pasj] {10.1093/pasj/psx078}, \href
  {https://ui.adsabs.harvard.edu/abs/2018PASJ...70S..21K} {70, S21}

\bibitem[\protect\citeauthoryear{{Lai} et~al.,}{{Lai}
  et~al.}{2023}]{LaiDing2023}
{Lai} L.~M.,  et~al., 2023, \mn@doi [arXiv e-prints]
  {10.48550/arXiv.2312.03244}, \href
  {https://ui.adsabs.harvard.edu/abs/2023arXiv231203244L} {p. arXiv:2312.03244}

\bibitem[\protect\citeauthoryear{Landy \& Szalay}{Landy \&
  Szalay}{1993}]{landy1993bias}
Landy S.~D.,  Szalay A.~S.,  1993, Astrophysical Journal, Part 1 (ISSN
  0004-637X), vol. 412, no. 1, p. 64-71., 412, 64

\bibitem[\protect\citeauthoryear{{Law-Smith} \& {Eisenstein}}{{Law-Smith} \&
  {Eisenstein}}{2017}]{LawSmith_Eisenstein_2017_xir_depend}
{Law-Smith} J.,  {Eisenstein} D.~J.,  2017, \mn@doi [\apj]
  {10.3847/1538-4357/836/1/87}, \href
  {https://ui.adsabs.harvard.edu/abs/2017ApJ...836...87L} {836, 87}

\bibitem[\protect\citeauthoryear{{Le F{\`e}vre} et~al.,}{{Le F{\`e}vre}
  et~al.}{2015}]{LeFevre+2025_VIMOS}
{Le F{\`e}vre} O.,  et~al., 2015, \mn@doi [\aap] {10.1051/0004-6361/201423829},
  \href {https://ui.adsabs.harvard.edu/abs/2015A&A...576A..79L} {576, A79}

\bibitem[\protect\citeauthoryear{Lee, Ferguson, Somerville, Wiklind  \&
  Giavalisco}{Lee et~al.}{2010}]{Lee+2010_SFH_uncertainty}
Lee S.-K.,  Ferguson H.~C.,  Somerville R.~S.,  Wiklind T.,   Giavalisco M.,
  2010, \mn@doi [The Astrophysical Journal] {10.1088/0004-637x/725/2/1644},
  725, 1644–1651

\bibitem[\protect\citeauthoryear{{Lemaux} et~al.,}{{Lemaux}
  et~al.}{2022}]{Lemaux+2022}
{Lemaux} B.~C.,  et~al., 2022, \mn@doi [\aap] {10.1051/0004-6361/202039346},
  \href {https://ui.adsabs.harvard.edu/abs/2022A&A...662A..33L} {662, A33}

\bibitem[\protect\citeauthoryear{Li et~al.,}{Li et~al.}{2022}]{li2022three}
Li X.,  et~al., 2022, Publications of the Astronomical Society of Japan, 74,
  421

\bibitem[\protect\citeauthoryear{{Limber}}{{Limber}}{1953}]{Limber1953}
{Limber} D.~N.,  1953, \mn@doi [\apj] {10.1086/145672}, \href
  {https://ui.adsabs.harvard.edu/abs/1953ApJ...117..134L} {117, 134}

\bibitem[\protect\citeauthoryear{{Lin}, {Fang}, {Cai}, {Wang}, {Fan}  \&
  {Kong}}{{Lin} et~al.}{2019}]{Lin_Fang+2019_color_xir_cosmos}
{Lin} X.,  {Fang} G.,  {Cai} Z.-Y.,  {Wang} T.,  {Fan} L.,   {Kong} X.,  2019,
  \mn@doi [\apj] {10.3847/1538-4357/ab0e73}, \href
  {https://ui.adsabs.harvard.edu/abs/2019ApJ...875...83L} {875, 83}

\bibitem[\protect\citeauthoryear{{Madau}, {Pozzetti}  \& {Dickinson}}{{Madau}
  et~al.}{1998}]{Madau+1998_sfh}
{Madau} P.,  {Pozzetti} L.,   {Dickinson} M.,  1998, \mn@doi [\apj]
  {10.1086/305523}, \href
  {https://ui.adsabs.harvard.edu/abs/1998ApJ...498..106M} {498, 106}

\bibitem[\protect\citeauthoryear{Mandelbaum et~al.,}{Mandelbaum
  et~al.}{2018}]{mandelbaum2018first}
Mandelbaum R.,  et~al., 2018, Publications of the Astronomical Society of
  Japan, 70, S25

\bibitem[\protect\citeauthoryear{{Mao}, {Kodama}, {P{\'e}rez-Mart{\'\i}nez},
  {Suzuki}, {Yamamoto}  \& {Adachi}}{{Mao}
  et~al.}{2022}]{Mao_Kodama+2022_galaxy_quench_sfr_env_z1_cosmos}
{Mao} Z.,  {Kodama} T.,  {P{\'e}rez-Mart{\'\i}nez} J.~M.,  {Suzuki} T.~L.,
  {Yamamoto} N.,   {Adachi} K.,  2022, \mn@doi [\aap]
  {10.1051/0004-6361/202243733}, \href
  {https://ui.adsabs.harvard.edu/abs/2022A&A...666A.141M} {666, A141}

\bibitem[\protect\citeauthoryear{Martinez, Arnalte-Mur  \& Stoyan}{Martinez
  et~al.}{2010}]{martinez2010measuring}
Martinez V.~J.,  Arnalte-Mur P.,   Stoyan D.,  2010, Astronomy \& Astrophysics,
  513, A22

\bibitem[\protect\citeauthoryear{{Massara}, {Villaescusa-Navarro}, {Ho},
  {Dalal}  \& {Spergel}}{{Massara} et~al.}{2021}]{Massara2021_mcf_neutrino}
{Massara} E.,  {Villaescusa-Navarro} F.,  {Ho} S.,  {Dalal} N.,   {Spergel}
  D.~N.,  2021, \mn@doi [\prl] {10.1103/PhysRevLett.126.011301}, \href
  {https://ui.adsabs.harvard.edu/abs/2021PhRvL.126a1301M} {126, 011301}

\bibitem[\protect\citeauthoryear{{Massara} et~al.,}{{Massara}
  et~al.}{2023}]{Massara2023_mcf_cosmology}
{Massara} E.,  et~al., 2023, \mn@doi [\apj] {10.3847/1538-4357/acd44d}, \href
  {https://ui.adsabs.harvard.edu/abs/2023ApJ...951...70M} {951, 70}

\bibitem[\protect\citeauthoryear{Mitchell, Lacey, Baugh  \& Cole}{Mitchell
  et~al.}{2013}]{Mitchell+2013_Mstar_sfh_SED}
Mitchell P.~D.,  Lacey C.~G.,  Baugh C.~M.,   Cole S.,  2013, Mon. Not. R.
  Astron. Soc., 435, 87

\bibitem[\protect\citeauthoryear{Miyazaki et~al.,}{Miyazaki
  et~al.}{2018}]{miyazaki2018hyper}
Miyazaki S.,  et~al., 2018, Publications of the Astronomical Society of Japan,
  70, S1

\bibitem[\protect\citeauthoryear{{Morales} et~al.,}{{Morales}
  et~al.}{2024}]{Morales+2024_UVbeta}
{Morales} A.,  et~al., 2024, \mn@doi [arXiv e-prints]
  {10.48550/arXiv.2405.20901}, \href
  {https://ui.adsabs.harvard.edu/abs/2024arXiv240520901M} {p. arXiv:2405.20901}

\bibitem[\protect\citeauthoryear{{Mostek}, {Coil}, {Cooper}, {Davis}, {Newman}
  \& {Weiner}}{{Mostek} et~al.}{2013}]{Mostek+2013_xir_depend_deep2}
{Mostek} N.,  {Coil} A.~L.,  {Cooper} M.,  {Davis} M.,  {Newman} J.~A.,
  {Weiner} B.~J.,  2013, \mn@doi [\apj] {10.1088/0004-637X/767/1/89}, \href
  {https://ui.adsabs.harvard.edu/abs/2013ApJ...767...89M} {767, 89}

\bibitem[\protect\citeauthoryear{{Muldrew} et~al.,}{{Muldrew}
  et~al.}{2012}]{muldrew_2012_gal_env}
{Muldrew} S.~I.,  et~al., 2012, \mn@doi [\mnras]
  {10.1111/j.1365-2966.2011.19922.x}, \href
  {https://ui.adsabs.harvard.edu/abs/2012MNRAS.419.2670M} {419, 2670}

\bibitem[\protect\citeauthoryear{Nishizawa, Hsieh, Tanaka  \& Takata}{Nishizawa
  et~al.}{2020}]{nishizawa2020photometric}
Nishizawa A.~J.,  Hsieh B.-C.,  Tanaka M.,   Takata T.,  2020, arXiv preprint
  arXiv:2003.01511

\bibitem[\protect\citeauthoryear{Norberg et~al.,}{Norberg
  et~al.}{2002}]{Norberg2002}
Norberg P.,  et~al., 2002, \mn@doi [Monthly Notices of the Royal Astronomical
  Society] {10.1046/j.1365-8711.2002.05348.x}, 332, 827

\bibitem[\protect\citeauthoryear{Norberg, Baugh, Gaztanaga  \& Croton}{Norberg
  et~al.}{2009}]{norberg2009statistical}
Norberg P.,  Baugh C.~M.,  Gaztanaga E.,   Croton D.~J.,  2009, Monthly Notices
  of the Royal Astronomical Society, 396, 19

\bibitem[\protect\citeauthoryear{{Old} et~al.,}{{Old}
  et~al.}{2020}]{old_balogh+2020}
{Old} L.~J.,  et~al., 2020, \mn@doi [\mnras] {10.1093/mnras/staa579}, \href
  {https://ui.adsabs.harvard.edu/abs/2020MNRAS.493.5987O} {493, 5987}

\bibitem[\protect\citeauthoryear{Ono et~al.,}{Ono et~al.}{2018}]{ono2018great}
Ono Y.,  et~al., 2018, Publications of the Astronomical Society of Japan, 70,
  S10

\bibitem[\protect\citeauthoryear{{Paranjape}, {Kova{\v{c}}}, {Hartley}  \&
  {Pahwa}}{{Paranjape} et~al.}{2015}]{paranjape+2015_conformity_conc}
{Paranjape} A.,  {Kova{\v{c}}} K.,  {Hartley} W.~G.,   {Pahwa} I.,  2015,
  \mn@doi [\mnras] {10.1093/mnras/stv2137}, \href
  {https://ui.adsabs.harvard.edu/abs/2015MNRAS.454.3030P} {454, 3030}

\bibitem[\protect\citeauthoryear{Peebles}{Peebles}{1980}]{Peebles1980}
Peebles P.~J.~E.,  1980, The large-scale structure of the universe

\bibitem[\protect\citeauthoryear{{Peng}, {Lilly}, {Renzini}  \&
  {Carollo}}{{Peng} et~al.}{2012}]{Peng+2012}
{Peng} Y.-j.,  {Lilly} S.~J.,  {Renzini} A.,   {Carollo} M.,  2012, \mn@doi
  [\apj] {10.1088/0004-637X/757/1/4}, \href
  {https://ui.adsabs.harvard.edu/abs/2012ApJ...757....4P} {757, 4}

\bibitem[\protect\citeauthoryear{{Planck Collaboration} et~al.,}{{Planck
  Collaboration} et~al.}{2020}]{Planck_2018}
{Planck Collaboration} et~al., 2020, \mn@doi [\aap]
  {10.1051/0004-6361/201833910}, \href
  {https://ui.adsabs.harvard.edu/abs/2020A&A...641A...6P} {641, A6}

\bibitem[\protect\citeauthoryear{{Riggs}, {Barbhuiyan}, {Loveday}, {Brough},
  {Holwerda}, {Hopkins}  \& {Phillipps}}{{Riggs} et~al.}{2021}]{Riggs2021}
{Riggs} S.~D.,  {Barbhuiyan} R.~W.~Y.~M.,  {Loveday} J.,  {Brough} S.,
  {Holwerda} B.~W.,  {Hopkins} A.~M.,   {Phillipps} S.,  2021, \mn@doi [\mnras]
  {10.1093/mnras/stab1697}, \href
  {https://ui.adsabs.harvard.edu/abs/2021MNRAS.506...21R} {506, 21}

\bibitem[\protect\citeauthoryear{Rutherford et~al.,}{Rutherford
  et~al.}{2021}]{rutherford2021sami}
Rutherford T.~H.,  et~al., 2021, The Astrophysical Journal, 918, 84

\bibitem[\protect\citeauthoryear{Satpathy, A~C~Croft, Ho  \& Li}{Satpathy
  et~al.}{2019}]{satpathy2019measurement}
Satpathy S.,  A~C~Croft R.,  Ho S.,   Li B.,  2019, Monthly Notices of the
  Royal Astronomical Society, 484, 2148

\bibitem[\protect\citeauthoryear{{Sawicki} et~al.,}{{Sawicki}
  et~al.}{2019}]{sawicki+2019_clauds}
{Sawicki} M.,  et~al., 2019, \mn@doi [\mnras] {10.1093/mnras/stz2522}, \href
  {https://ui.adsabs.harvard.edu/abs/2019MNRAS.489.5202S} {489, 5202}

\bibitem[\protect\citeauthoryear{Schaerer, de Barros  \& Sklias}{Schaerer
  et~al.}{2013}]{Schaerer+2013_LBG_sfh_SED}
Schaerer D.,  de Barros S.,   Sklias P.,  2013, Astron. Astrophys., 549, A4

\bibitem[\protect\citeauthoryear{Scranton et~al.,}{Scranton
  et~al.}{2002}]{scranton2002analysis}
Scranton R.,  et~al., 2002, The Astrophysical Journal, 579, 48

\bibitem[\protect\citeauthoryear{{Sheth}}{{Sheth}}{2005}]{Sheth2005_marked}
{Sheth} R.~K.,  2005, \mn@doi [\mnras] {10.1111/j.1365-2966.2005.09609.x},
  \href {https://ui.adsabs.harvard.edu/abs/2005MNRAS.364..796S} {364, 796}

\bibitem[\protect\citeauthoryear{Sheth \& Tormen}{Sheth \&
  Tormen}{2004a}]{sheth2004environmental}
Sheth R.~K.,  Tormen G.,  2004a, Monthly Notices of the Royal Astronomical
  Society, 350, 1385

\bibitem[\protect\citeauthoryear{{Sheth} \& {Tormen}}{{Sheth} \&
  {Tormen}}{2004b}]{Sheth+2004_marked}
{Sheth} R.~K.,  {Tormen} G.,  2004b, \mn@doi [\mnras]
  {10.1111/j.1365-2966.2004.07733.x}, \href
  {https://ui.adsabs.harvard.edu/abs/2004MNRAS.350.1385S} {350, 1385}

\bibitem[\protect\citeauthoryear{Sheth, Connolly  \& Skibba}{Sheth
  et~al.}{2005}]{sheth_connolly_2005MCF}
Sheth R.~K.,  Connolly A.~J.,   Skibba R.,  2005, arXiv preprint
  astro-ph/0511773

\bibitem[\protect\citeauthoryear{Sheth, Jimenez, Panter  \& Heavens}{Sheth
  et~al.}{2006a}]{sheth2006environment}
Sheth R.~K.,  Jimenez R.,  Panter B.,   Heavens A.~F.,  2006a, The
  Astrophysical Journal, 650, L25

\bibitem[\protect\citeauthoryear{{Sheth}, {Jimenez}, {Panter}  \&
  {Heavens}}{{Sheth} et~al.}{2006b}]{Sheth+2006_marked}
{Sheth} R.~K.,  {Jimenez} R.,  {Panter} B.,   {Heavens} A.~F.,  2006b, \mn@doi
  [\apjl] {10.1086/508683}, \href
  {https://ui.adsabs.harvard.edu/abs/2006ApJ...650L..25S} {650, L25}

\bibitem[\protect\citeauthoryear{{Shi}, {Malavasi}, {Toshikawa}  \&
  {Zheng}}{{Shi} et~al.}{2024}]{shi+2024_sfr_env_z4}
{Shi} K.,  {Malavasi} N.,  {Toshikawa} J.,   {Zheng} X.,  2024, \mn@doi [\apj]
  {10.3847/1538-4357/ad11d7}, \href
  {https://ui.adsabs.harvard.edu/abs/2024ApJ...961...39S} {961, 39}

\bibitem[\protect\citeauthoryear{{Shibuya}, {Ouchi}  \& {Harikane}}{{Shibuya}
  et~al.}{2015}]{shibuya+2015_muv_mstar}
{Shibuya} T.,  {Ouchi} M.,   {Harikane} Y.,  2015, \mn@doi [\apjs]
  {10.1088/0067-0049/219/2/15}, \href
  {https://ui.adsabs.harvard.edu/abs/2015ApJS..219...15S} {219, 15}

\bibitem[\protect\citeauthoryear{{Shimakawa} et~al.,}{{Shimakawa}
  et~al.}{2018}]{shimakawa+2018_sfr_env}
{Shimakawa} R.,  et~al., 2018, \mn@doi [\mnras] {10.1093/mnras/stx2494}, \href
  {https://ui.adsabs.harvard.edu/abs/2018MNRAS.473.1977S} {473, 1977}

\bibitem[\protect\citeauthoryear{Skibba, Sheth, Connolly  \& Scranton}{Skibba
  et~al.}{2006}]{skibba2006MCF}
Skibba R.,  Sheth R.~K.,  Connolly A.~J.,   Scranton R.,  2006, Monthly Notices
  of the Royal Astronomical Society, 369, 68

\bibitem[\protect\citeauthoryear{Skibba et~al.,}{Skibba
  et~al.}{2012}]{skibba2012galaxy}
Skibba R.~A.,  et~al., 2012, Monthly Notices of the Royal Astronomical Society,
  423, 1485

\bibitem[\protect\citeauthoryear{Skibba, Sheth, Croton, Muldrew, Abbas, Pearce
  \& Shattow}{Skibba et~al.}{2013}]{skibba2013measures}
Skibba R.~A.,  Sheth R.~K.,  Croton D.~J.,  Muldrew S.~I.,  Abbas U.,  Pearce
  F.~R.,   Shattow G.~M.,  2013, Monthly Notices of the Royal Astronomical
  Society, 429, 458

\bibitem[\protect\citeauthoryear{{Sobral}, {Best}, {Smail}, {Geach},
  {Cirasuolo}, {Garn}  \& {Dalton}}{{Sobral}
  et~al.}{2011}]{Sobral+2011_sfr_env_Mstar}
{Sobral} D.,  {Best} P.~N.,  {Smail} I.,  {Geach} J.~E.,  {Cirasuolo} M.,
  {Garn} T.,   {Dalton} G.~B.,  2011, \mn@doi [\mnras]
  {10.1111/j.1365-2966.2010.17707.x}, \href
  {https://ui.adsabs.harvard.edu/abs/2011MNRAS.411..675S} {411, 675}

\bibitem[\protect\citeauthoryear{{Song} et~al.,}{{Song}
  et~al.}{2016}]{song+2016_Muv_Mstar}
{Song} M.,  et~al., 2016, \mn@doi [\apj] {10.3847/0004-637X/825/1/5}, \href
  {https://ui.adsabs.harvard.edu/abs/2016ApJ...825....5S} {825, 5}

\bibitem[\protect\citeauthoryear{Steidel, Giavalisco, Pettini, Dickinson  \&
  Adelberger}{Steidel et~al.}{1996}]{steidel1996spectroscopic}
Steidel C.~C.,  Giavalisco M.,  Pettini M.,  Dickinson M.,   Adelberger K.~L.,
  1996, The Astrophysical Journal, 462, L17

\bibitem[\protect\citeauthoryear{{Storey-Fisher}, {Tinker}, {Zhai}, {DeRose},
  {Wechsler}  \& {Banerjee}}{{Storey-Fisher} et~al.}{2022}]{SFisher+_2022}
{Storey-Fisher} K.,  {Tinker} J.,  {Zhai} Z.,  {DeRose} J.,  {Wechsler} R.~H.,
   {Banerjee} A.,  2022, \mn@doi [arXiv e-prints] {10.48550/arXiv.2210.03203},
  \href {https://ui.adsabs.harvard.edu/abs/2022arXiv221003203S} {p.
  arXiv:2210.03203}

\bibitem[\protect\citeauthoryear{Sureshkumar et~al.,}{Sureshkumar
  et~al.}{2021}]{sureshkumar2021_gamaMCF}
Sureshkumar U.,  et~al., 2021, Astronomy \& Astrophysics, 653, A35

\bibitem[\protect\citeauthoryear{Sureshkumar et~al.,}{Sureshkumar
  et~al.}{2023a}]{sureshkumar2023_midinfra}
Sureshkumar U.,  et~al., 2023a, Astronomy \& Astrophysics, 669, A27

\bibitem[\protect\citeauthoryear{{Sureshkumar} et~al.,}{{Sureshkumar}
  et~al.}{2023b}]{sureshkumar2023}
{Sureshkumar} U.,  et~al., 2023b, \mn@doi [\aap] {10.1051/0004-6361/202243193},
  \href {https://ui.adsabs.harvard.edu/abs/2023A&A...669A..27S} {669, A27}

\bibitem[\protect\citeauthoryear{{Sureshkumar} et~al.,}{{Sureshkumar}
  et~al.}{2024}]{sureshkumar2024_merger}
{Sureshkumar} U.,  et~al., 2024, \mn@doi [\aap] {10.1051/0004-6361/202347705},
  \href {https://ui.adsabs.harvard.edu/abs/2024A&A...686A..40S} {686, A40}

\bibitem[\protect\citeauthoryear{{Szewciw}, {Beltz-Mohrmann}, {Berlind}  \&
  {Sinha}}{{Szewciw} et~al.}{2022}]{szewciw+2022}
{Szewciw} A.~O.,  {Beltz-Mohrmann} G.~D.,  {Berlind} A.~A.,   {Sinha} M.,
  2022, \mn@doi [\apj] {10.3847/1538-4357/ac3a7c}, \href
  {https://ui.adsabs.harvard.edu/abs/2022ApJ...926...15S} {926, 15}

\bibitem[\protect\citeauthoryear{{Taamoli} et~al.,}{{Taamoli}
  et~al.}{2023}]{Taamoli+2023}
{Taamoli} S.,  et~al., 2023, \mn@doi [arXiv e-prints]
  {10.48550/arXiv.2312.10222}, \href
  {https://ui.adsabs.harvard.edu/abs/2023arXiv231210222T} {p. arXiv:2312.10222}

\bibitem[\protect\citeauthoryear{Tanaka et~al.,}{Tanaka
  et~al.}{2018}]{tanaka2018photometric}
Tanaka M.,  et~al., 2018, Publications of the Astronomical Society of Japan,
  70, S9

\bibitem[\protect\citeauthoryear{{Tinker}, {Robertson}, {Kravtsov}, {Klypin},
  {Warren}, {Yepes}  \& {Gottl{\"o}ber}}{{Tinker}
  et~al.}{2010}]{Tinker+2010_bias}
{Tinker} J.~L.,  {Robertson} B.~E.,  {Kravtsov} A.~V.,  {Klypin} A.,  {Warren}
  M.~S.,  {Yepes} G.,   {Gottl{\"o}ber} S.,  2010, \mn@doi [\apj]
  {10.1088/0004-637X/724/2/878}, \href
  {https://ui.adsabs.harvard.edu/abs/2010ApJ...724..878T} {724, 878}

\bibitem[\protect\citeauthoryear{Toshikawa et~al.,}{Toshikawa
  et~al.}{2018}]{toshikawa2018goldrush}
Toshikawa J.,  et~al., 2018, Publications of the Astronomical Society of Japan,
  70, S12

\bibitem[\protect\citeauthoryear{{Toshikawa} et~al.,}{{Toshikawa}
  et~al.}{2024}]{toshikawa+2024}
{Toshikawa} J.,  et~al., 2024, \mn@doi [\mnras] {10.1093/mnras/stad3162}, \href
  {https://ui.adsabs.harvard.edu/abs/2024MNRAS.527.6276T} {527, 6276}

\bibitem[\protect\citeauthoryear{{Weaver} et~al.,}{{Weaver}
  et~al.}{2022}]{weaver+2022_cosmos}
{Weaver} J.~R.,  et~al., 2022, \mn@doi [\apjs] {10.3847/1538-4365/ac3078},
  \href {https://ui.adsabs.harvard.edu/abs/2022ApJS..258...11W} {258, 11}

\bibitem[\protect\citeauthoryear{{White}}{{White}}{2016}]{white2016_mcf_modgrav}
{White} M.,  2016, \mn@doi [\jcap] {10.1088/1475-7516/2016/11/057}, \href
  {https://ui.adsabs.harvard.edu/abs/2016JCAP...11..057W} {2016, 057}

\bibitem[\protect\citeauthoryear{White \& Padmanabhan}{White \&
  Padmanabhan}{2009}]{white2009breaking}
White M.,  Padmanabhan N.,  2009, Monthly Notices of the Royal Astronomical
  Society, 395, 2381

\bibitem[\protect\citeauthoryear{Wilkins, Gonzalez-Perez, Lacey  \&
  Baugh}{Wilkins et~al.}{2012}]{Wilkins+2012_UV_SFR_relation}
Wilkins S.~M.,  Gonzalez-Perez V.,  Lacey C.~G.,   Baugh C.~M.,  2012, Mon.
  Not. R. Astron. Soc., 427, 1490

\bibitem[\protect\citeauthoryear{{Woo} et~al.,}{{Woo}
  et~al.}{2013}]{woo_dekel+2013}
{Woo} J.,  et~al., 2013, \mn@doi [\mnras] {10.1093/mnras/sts274}, \href
  {https://ui.adsabs.harvard.edu/abs/2013MNRAS.428.3306W} {428, 3306}

\bibitem[\protect\citeauthoryear{{Xiao} et~al.,}{{Xiao}
  et~al.}{2022a}]{Xiao2022}
{Xiao} X.,  et~al., 2022a, \mn@doi [\mnras] {10.1093/mnras/stac879}, \href
  {https://ui.adsabs.harvard.edu/abs/2022MNRAS.513..595X} {513, 595}

\bibitem[\protect\citeauthoryear{{Xiao} et~al.,}{{Xiao}
  et~al.}{2022b}]{Xiao2022_mcf_cosmology}
{Xiao} X.,  et~al., 2022b, \mn@doi [\mnras] {10.1093/mnras/stac879}, \href
  {https://ui.adsabs.harvard.edu/abs/2022MNRAS.513..595X} {513, 595}

\bibitem[\protect\citeauthoryear{Yang et~al.,}{Yang
  et~al.}{2020}]{Yang_2020_cosmology}
Yang Y.,  et~al., 2020, \mn@doi [The Astrophysical Journal]
  {10.3847/1538-4357/aba35b}, 900, 6

\bibitem[\protect\citeauthoryear{Zehavi et~al.,}{Zehavi
  et~al.}{2005}]{Zehavi2005}
Zehavi I.,  et~al., 2005, \mn@doi [The Astrophysical Journal] {10.1086/431891},
  630, 1

\bibitem[\protect\citeauthoryear{{Zehavi} et~al.,}{{Zehavi}
  et~al.}{2011}]{zehavi+2011}
{Zehavi} I.,  et~al., 2011, \mn@doi [\apj] {10.1088/0004-637X/736/1/59}, \href
  {https://ui.adsabs.harvard.edu/abs/2011ApJ...736...59Z} {736, 59}

\bibitem[\protect\citeauthoryear{{Zhai}, {Percival}  \& {Guo}}{{Zhai}
  et~al.}{2023}]{Zhai_Percival_2023_xir_depend_boss}
{Zhai} Z.,  {Percival} W.~J.,   {Guo} H.,  2023, \mn@doi [\mnras]
  {10.1093/mnras/stad1793}, \href
  {https://ui.adsabs.harvard.edu/abs/2023MNRAS.523.5538Z} {523, 5538}

\bibitem[\protect\citeauthoryear{{Zhang} et~al.,}{{Zhang}
  et~al.}{2022}]{zhang+2022_submm-highz-gals}
{Zhang} Y.,  et~al., 2022, \mn@doi [\mnras] {10.1093/mnras/stac824}, \href
  {https://ui.adsabs.harvard.edu/abs/2022MNRAS.512.4893Z} {512, 4893}

\bibitem[\protect\citeauthoryear{{Zu} \& {Mandelbaum}}{{Zu} \&
  {Mandelbaum}}{2018}]{zu_2018_conformity_mcf}
{Zu} Y.,  {Mandelbaum} R.,  2018, \mn@doi [\mnras] {10.1093/mnras/sty279},
  \href {https://ui.adsabs.harvard.edu/abs/2018MNRAS.476.1637Z} {476, 1637}

\makeatother
\end{thebibliography}
\end{document}